\documentclass[12pt,preprint]{aastex}
\usepackage{enumerate}
\usepackage{float}
\usepackage{placeins}
\usepackage{amsmath}
\usepackage{natbib}
\usepackage{epstopdf}
\usepackage{lscape}
\usepackage{ctable} 
\usepackage{color}
\usepackage[utf8x]{inputenx}
\usepackage[english]{babel}
\usepackage{footnote}
\usepackage{pdflscape}
\usepackage{verbatim}
\usepackage{rotating}
\newcommand{\HI}{H\,{\sc i}~}
\newcommand{\HII}{H\,{\sc ii}~}

\defcitealias{2014ApJ...786..155N}{Paper I}
\begin{document}

\title{Metal-poor dwarf galaxies in the SIGRID galaxy sample. II. The electron temperature--abundance calibration and the parameters that affect it}
\shorttitle{Low metallicity SIGRID galaxies}
\shortauthors{Nicholls et~al.}

\author{ David C. Nicholls\altaffilmark{1}, Michael A. Dopita\altaffilmark{1}$^,$\altaffilmark{2},  Ralph S. Sutherland\altaffilmark{1}, Helmut Jerjen\altaffilmark{1}, Lisa J. Kewley\altaffilmark{1}, \& Hassan Basurah\altaffilmark{2},}
\email{David.Nicholls@anu.edu.au}
\altaffiltext{1}{Research School of Astronomy and Astrophysics, Australian National University, Cotter Rd., Weston ACT 2611, Australia }
\altaffiltext{2}{Astronomy Department, King Abdulaziz University, P.O. Box 80203, Jeddah, Saudi Arabia}

\begin{abstract}
In this paper, we use the Mappings photoionization code to explore the physical parameters that impact on the measurement of electron temperature and abundance in \HII regions. In the previous paper we presented observations and measurements of physical properties from the spectra of seventeen \HII regions in fourteen isolated dwarf irregular galaxies from the SIGRID sample. Here, we analyze these observations further, together with three additional published data sets. We explore the effects of optical thickness, electron density, ionization parameter, ionization source, and non-equilibrium effects on the relation between electron temperature and metallicity.  We present a standard model that fits the observed data remarkably well at metallicities between 1/10  and 1 solar. We investigate the effects of optically thin \HII regions, and show that they can have a considerable effect on the measured electron temperature, and that there is evidence that some of the observed objects are optically thin. We look at the role of the ionization parameter and find that lower ionization parameter values give better fits at higher oxygen abundance. We show that higher pressures combined with low optical depth, and also $\kappa$ electron energy distributions at low $\kappa$ values, can generate the apparent high electron temperatures in low metallicity \HII regions, and that the former provides the better fit to observations.  We examine the effects of these parameters on the strong line diagnostic methods. We extend this to three-dimensional diagnostic grids to confirm how well the observations are described by the grids.

\end{abstract}


\section{Introduction}

The term ``electron temperature'' in an \HII region usually refers to the apparent [O~{\sc iii}] electron temperature, T$_e$, derived from the [O~{\sc iii}] optical spectra of the nebula, using a simple method based on the relative collisional excitation rates of the $^1$D$_2$ and $^1$S$_0$ levels of O$^{++}$ \citep[see, e.g.,][]{2006agna.book.....O, 2012ApJ...752..148N}. This (with similar measurements for O$^+$) is used as the basis for calculating the total oxygen abundance, when the electron temperatures are available.

Spectra arising from collisional excitation in a particular region in the nebula will exhibit the characteristics of the local electron temperature (or electron energy distribution) and radiative environment of that region, but observed electron temperatures are only mean electron temperatures, as they are calculated from emission lines averaged over a range of zones in the nebula, and with a range of physical temperatures. The abundance of oxygen is then determined from the integrated fluxes of [O~{\sc iii}] and [O~{\sc ii}], using the derived [O~{\sc iii}] and [O~{\sc ii}] electron temperatures \citep[see, e.g.,][]{2006A&A...448..955I}. The apparent abundance is, therefore, only approximately related to the true oxygen abundance. The matter is complicated further if there are non-equilibrium electron energy distributions, as suggested by \cite{2012ApJ...752..148N, 2013ApJS..207...21N} and \cite{2013ApJS..208...10D}.

Due to the highly non-uniform physical structures in real \HII regions, no general photoionization model is yet capable of precisely reproducing the observed spectral emissions. The simplest approximation to an \HII region, is the uniform ``single slab'' model. This is a poor approximation to the behavior of a real \HII region, as it assumes all the ionic species are uniformly distributed and at the same physical temperature, and needs to be corrected to reflect the true distribution of ions and temperatures. For example, the [O~{\sc ii}] emission arises predominantly from the cooler (usually outer) regions of the nebula  whereas the [O~{\sc iii}] emission is more uniformly distributed. Assuming that the two ions are emitting at the same temperature will overestimate the [O~{\sc ii}] flux (per atom) and thus underestimate the [O~{\sc ii}] abundance in a real nebula.  Empirical or theory-based correction factors are often applied \citep{1992AJ....103.1330G,2012MNRAS.426.2630L}. Other ionization states, for example, O$^{+++}$ and O, occur in the highest and lowest excitation (usually innermost and outmost) regions of the nebula, respectively.  Unobserved species such as O$^{+++}$ which affect the populations of observed species are accounted for the so-called ``ionization correction factors'', discussed by many authors \citep[see, for example,][]{1969BOTT....5....3P,1978A&A....66..257S,1984ASSL..112.....A,1987MNRAS.226...19D,1994ApJ...435..647I,1995A&A...300...78E, 1996MNRAS.280..720V, 2004MNRAS.355..229E,2005A&A...434..507S,2005A&A...441..981B,2006A&A...448..955I,2009A&A...508..615L}.

Taking into account this three-dimensional structure, a more realistic model is the three dimensional ``Str\"omgren sphere'', with a stellar or star cluster  excitation source at the centre, a stellar-wind-evacuated central region, and radial zones of ionized gas at different physical temperatures.  These give rise to differing amounts of spectral emission from different ionic species, the observed results of which are the integrated spectral emissions from all the radial zones for each ionic species. The models can then be compared with the spectra observed in real \HII regions.   Current photoionization models such as Mappings IV \citep{2013ApJS..208...10D} are based on this physical structure. 

A still better approximation to real nebulae is to assume a spatially inhomogeneous turbulent structure for the nebula \citep[Sutherland et al., in progress, and earlier work,][]{2011EAS....48..397B}, but modeling this is significantly more computationally demanding. Indeed, one look at the complexities of the Orion, M17 and 30-Doradus nebulae shows such variation in form as to defy accurate modeling.  While extra-galactic \HII regions appear unresolved in more distant galaxies, they are very likely structurally as complex as the nearby nebulae, and at best, even inhomogeneous and fractal models are only approximations to reality. Nonetheless, spherical models are very useful in exploring the effects of different physical parameters on the observed spectra.

In this work we use the Str\"omgren Sphere model, using the Mappings IV photoionization code \citep[see appendix for the details of changes in the current version of the code since that described in][]{2013ApJS..208...10D}\footnote{Mappings is a comprehensive modeling code capable of studying a wide range of possible astrophysical nebular structures and physical phenomena, rather than a tailor-made model for specific objects. As such it may be compared with other similar complex modeling codes, as described by \cite{2001ASPC..247..533P}.}. To obtain information on the oxygen abundance of an observed but spatially unresolved \HII region, we need to identify the key physical parameters affecting the apparent electron temperature, and use these in the models to compare with observations.  From these we derive a relation between the intrinsic nebular oxygen abundance and the apparent (i.e., volume and luminosity averaged) [O~{\sc iii}] mean electron temperature. The models are then used to generate relationships between electron temperature, T$_e$, and oxygen abundance, $12+\log(O/H)$, for the model \HII region, to which the observed apparent electron temperature of a real object may compared.

The electron temperatures that our models generate within an \HII region vary between $\sim$30,000K and $\sim$100K, with 10,000K being typical. The inner zones (at higher temperatures) have smaller volumes, per unit thickness, than the outer zones.  Measurements of the [O~{\sc iii}] and [O~{\sc ii}] apparent electron temperatures from spectra reflect the average conditions in core and outer regions of the nebula, respectively.  The [O~{\sc iii}] temperature provides an average over the bulk of the internal regions of an \HII region, and is the most useful (and most widely used) single measurement of the conditions in the nebula. However, it is important to recognize that this single value does not characterize the entire \HII region.

This work continues on from our previous paper \citep[][hereafter ``paper 1'']{2014ApJ...786..155N}. The paper is structured as follows: in Section 2 we discuss the measurement of electron temperatures and the calculation of oxygen abundances, relating the outputs from the Mappings photoionization model code to data from the SIGRID sample \citep{2011AJ....142...83N} and \citetalias{2014ApJ...786..155N}, Sloan Digital Sky Survey (SDSS) data from \cite{2006A&A...448..955I} \citep[Data Release 3 from][]{2000AJ....120.1579Y}, low metallicity emission line galaxy data from \cite{2012A&A...546A.122I}, and data for Wolf-Rayet galaxies from \citet{2009A&A...508..615L,2010A&A...517A..85L}. These observations span a range of nebular metallicities between 1/50 and 1 solar.  We examine the calculation of the electron temperature and the chemical abundances.  In Section 3 we consider the electron temperature--chemical abundance relation. In Section 4 we examine the effect of optical depth at the Lyman edge of hydrogen at 912\AA.  In Section 5 we consider the effect of the  FUV spectral energy distribution of the central star cluster. In Section 6 we investigate the effects of pressure in \HII regions. In Section 7 we explore the role of the ionization parameter. In Section 8 we look at the effects dust. In Section 9 we explore the consequences of $\kappa$ non-equilibrium electron energy distributions. In Section 10 we summarize the effects of these parameters.  In Section 11 we look at the use of diagnostic line ratio grids from \cite{2013ApJS..208...10D}, and in Section 12 we extend this to examine the extra information yielded by using three-dimensional grids. In Section 13 we discuss our findings, and in Section 14 we present our conclusions.

\section{Electron temperature and oxygen abundance}
\subsection{Calculating electron temperatures}
Determining the electron temperature from [O~{\sc iii}] spectra from an \HII region has been discussed many times in the literature.  The simplest technique is to use a graph of the 4363\AA~and 5007+4959\AA~ [O~{\sc iii}] lines, as shown in  \citet[][Figure 5.1]{2006agna.book.....O}.  Other techniques have been proposed, such as the iterative method in \cite{2006A&A...448..955I}, which has its origin in work by \citet{1975MNRAS.170..475S}.  Here we have used the formulae derived from our Mappings photoionization models, presented in \cite{2013ApJS..207...21N}, using the O$^{++}$ collision strength data from \cite{2012MNRAS.423L..35P} (see Equation \ref{eq2}, below).  All methods are based on the change of the relative populations of the $^1$D$_2$ and $^1$S$_0$ energy levels of O$^{++}$ with temperature. When using the same O$^{++}$ collision strength data, the methods give very similar results. The choice of collision strength data is important, as different data can lead to electron temperatures differing by as much as 5\% at an electron temperature of 15,000K \citep[see Figure 2 in][]{2013ApJS..207...21N}. The Palay data tend to give lower [O~{\sc iii}] electron temperatures than older data sets. {The recent collision strength calculations by \citet{2013arXiv1311.6517S} tend to give higher electron temperatures, similar to those from \citet{1994A&AS..103..273L}. We have not been able to use these recent data in the Mappings photoionization models as calculations for the upper level data needed by Mappings do not converge (Storey, P, pers. comm.). For a detailed discussion of the effects of different collision strength data, see \citet{2013ApJS..207...21N}.

\subsection{Limitations in electron temperature calculations}

There is an important limitation in the methods used to measure the [O~{\sc iii}] electron temperature, for very low metallicity \HII regions. For the observational data considered in this paper, we have used the equations presented in \cite{2013ApJS..207...21N}. These are derived from the Mappings models in \cite{2013ApJS..208...10D}, and apply for metallicities between 1/20 and 1 solar.  Alternative methods used by \citet[][and references therein]{2006A&A...448..955I,2012A&A...546A.122I} have a longer history, and are based on the original work of \cite{1975MNRAS.170..475S}, for which fit parameters were only calculated  for electron temperatures between 5,000 and 20,000K.  The range restrictions in these methods limit the temperature ranges that can be calculated reliably. Although both methods can generate values for the electron temperature $>$ 20,000K, such values are the result of extrapolation outside the range of application, and are therefore not reliably based in physics.  At higher temperatures the physics coming into play involves processes that do not affect  the results at lower temperatures.  Models that do not allow for the increasing importance of the higher temperature processes will necessarily be unreliable. Any significant, unaccounted for, change in behavior outside the specified ranges will lead to erroneous calculated temperatures. Consequently, extrapolation is problematic, and values of T$_e$ $\gtrsim$ 20,000K cannot be considered reliable, if calculated using the methods discussed here.

\subsection{Oxygen abundance}
Oxygen abundance is used as a proxy for total metallicity as it is the dominant heavy element species observable in the optical spectrum, its spectral lines are present throughout \HII regions, and it is produced in stellar nuclear reactions that give rise to the other major heavy element components (e.g., C and N). It is present in \HII regions in five forms: O$^{+++}$, O$^{++}$, O$^{+}$, O$^0$ and in dust. O$^{+++}$ is present only in the highest ionization zones closest to the stellar excitation source in \HII regions, and does not contribute significantly to total oxygen. O$^{++}$ is the dominant form of oxygen in many low metallicity \HII regions where it occupies much of the body of the \HII region. O$^{+}$ is excited in the outer zones, or regions where the higher energy photons have been depleted, and O$^0$ is present only at the low-ionization outer edges.  However, some \HII regions have dominant O$^{+}$ excitation \citep[see, for example,][]{2009ApJ...700..654E, 2012A&A...546A.122I}. In the models presented here, we consider only a simple spherically symmetric model with a central strongly ionizing star cluster.

Figure \ref{fraco} illustrates these points, showing the ionic fractions for the gaseous oxygen species calculated using the Mappings IV photoionization code, as a function of radius for a spherical \HII region, from the minimum radius, R$_{min}$, at which the Str\"omgren sphere is gas-filled, to the maximum, R$_{max}$, defined as the radius at which the atomic hydrogen is 99\% neutral. The parameters for this calculation are $\log(q)$=7.5, 12+log(O/H)$_{total}$=0.1solar, optically thick, equilibrium electron energies, isobaric with $\log(P/k)$ = 5, and using the Starburst99 excitation model as described in Section 3, below. It is clear that O$^{++}$ is the dominant ionic fraction over most of the volume of the nebula, O$^+$ contributes mainly in the outer regions, and O$^0$ only at the outer edge or lowest excitation regions.

The ionization paramater, and the metallicity and oxygen abundance were defined in \citetalias{2014ApJ...786..155N}, but it is useful to repeat these definitions here. The ionization parameter $q$ (sometimes expressed as $\mathcal{U} = q/c$, where $c$ is the speed of light) is the ratio of the number of ionizing source photons passing through a unit volume to the neutral hydrogen density.  The photon flux matches the number of new ions it produces, and as $q$ has the dimensions of velocity, it can be understood as the maximum speed at which the boundary of the ionized region can move outwards \citep{2003adu..book.....D}. $q$ is at its maximum at the inner edge of the ionized region of an \HII region, and falls to zero at the outer edge of the ionized nebula, where (if) the ionizing flux is fully depleted. A value of $\log(q)$ = 8.5 (corresponding to $\log(\mathcal{U})$ = -1.98)  is at the high end of values likely to be found in an \HII region \citep[see, for example,][]{2012ApJ...757..108Y}. A more typical value for \HII regions is $\log(q)$ = 7.5, and we use these values as reference levels in this paper. In these models, the density is set by the pressure and temperature, through the parameter $\log(P/k)$, discussed in detail in Section 6, below. As a standard, we use $\log(P/k)$ = 5.0. The local value of the electron density set in this way depends on both the $\log(P/k)$ value and the local electron temperature: the actual electron density is $n_e \sim$ 5 cm$^{-3}$.

The metallicity is usually expressed in terms of the total oxygen abundance, as oxygen is the most prominent heavy element measurable in the optical spectrum, and is the most numerous species in \HII regions after hydrogen and helium. It is defined in terms of numerical values as 12+log(O/H) .  An alternative method is to define metallicity in terms of solar abundances.  Here (and in the Mappings models) we use the solar data from \cite{2010Ap&SS.328..179G}. Metallicity is also sometimes described using the parameter Z, but this is also used as an abundance by mass, so we have not used it in this paper.  A further point to note is that the oxygen abundance measured from optical spectra is the gas-phase abundance, and does not account for oxygen locked up in dust.  In our models, we have assumed solar depletion levels in dust, resulting in a difference of 0.07 dex between total oxygen abundance and gas-phase abundance. See the detailed discussion in Section 8, below.

\begin{figure}[htbp]
\centering
\includegraphics[width=0.8\hsize]{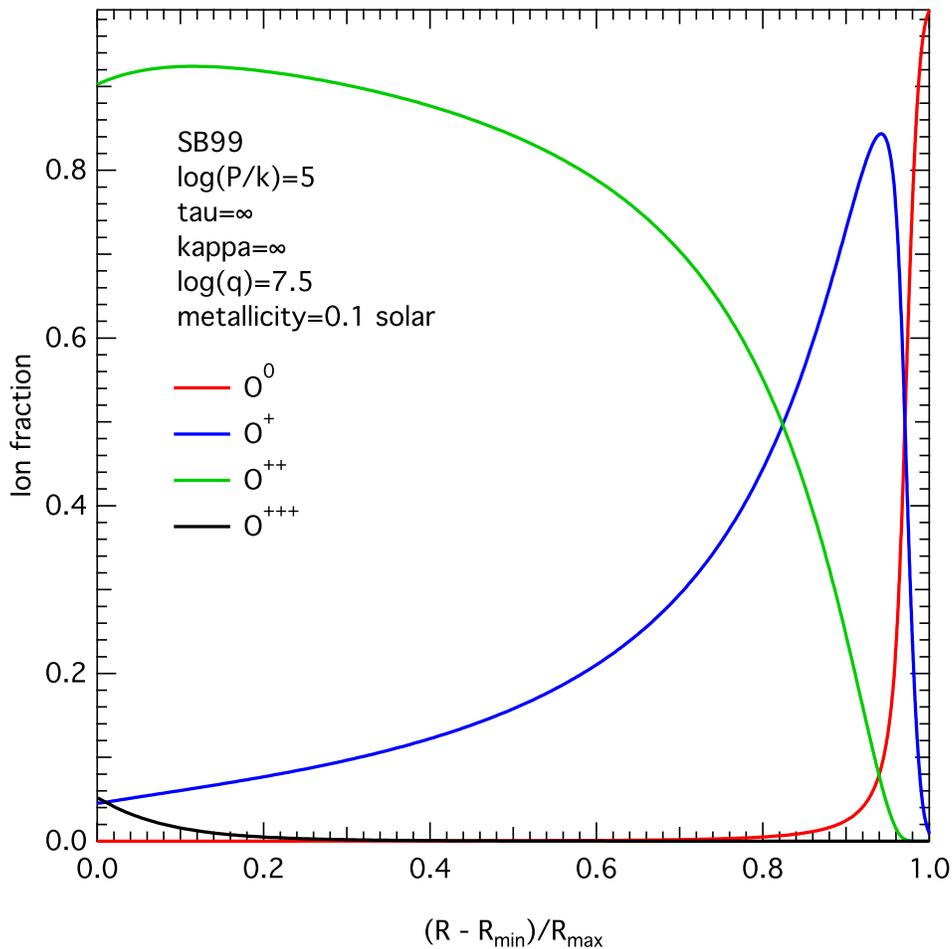}
\caption{Ionic fractions for oxygen species versus radius, from Mappings IV}\label{fraco}
\end{figure}

\FloatBarrier

\subsection{Calculating the oxygen abundance}
It is conventional to measure the apparent electron temperatures for O~{\sc iii} from the 4363, 4958 and 5007\AA~ lines, and for and O~{\sc ii} using the 3726/9 and 7320\AA~ region lines.  In the observations reported in \citetalias{2014ApJ...786..155N}, however, we did not measure the NIR [O~{\sc ii}] lines, nor the upper state [N~{\sc ii}] lines which provide an alternative measure, so we have calculated T$_e$(O~{\sc ii}) iteratively instead. A similar approach was taken by \cite{2006A&A...448..955I} for some of their data, and by \cite{2012MNRAS.426.2630L}.

As we showed in \citetalias{2014ApJ...786..155N}, the number density of O$^{++}$ ions to hydrogen ions (i.e., the O$^{++}$ abundance) can be expressed in terms of the flux ratio of O~{\sc iii}($^1$D$_2$) to H$_\beta$ \footnote{ $T_e$ is the electron temperature, $g_1$ is the statistical weight of the transition ground state, $k$ is the Boltzmann constant, $\Upsilon_{12}$ is the net effective collision strength for collisional excitations from the ground states to the upper state, E$_{12}$ is the energy level of the  upper state, and $\alpha^{eff}_B (\text{H}_{\beta})$ is the effective emissivity for H$\beta$.},
\begin{equation}\label{eq1}
\frac{n_{O^{++}}}{n_{H^+}} = \frac{\text{flux}(\text{O{\scshape iii})}}{\text{flux}(\text{H}_\beta)}.g_1.\sqrt{T_e}.\alpha^{eff}_B (\text{H}_{\beta}).\text{exp}(E_{12}/(kT_e)) \times115885.4/(E_{12}.\Upsilon_{12})
\end{equation}
where $T_e$ is the apparent electron temperature derived from the {O~{\sc iii}} line ratio, for which there is a simple expression from \cite{2013ApJS..207...21N},
\begin{equation}\label{eq2}
T_e = a~(-log_{10}({\mathcal{R}}) -b)^{-c} ,
\end{equation}
where the flux ratio for [O~{\sc iii}] is,
\begin{equation}\label{eq3}
\mathcal{R} = \frac{j(\lambda4363)}{j(\lambda5007)+j(\lambda4959)} , 
\end{equation}
and a= 13205, b= 0.92506, and c=0.98062. These parameters have been revised slightly from the earlier version, based on the latest outputs from the Mappings IV photoionization code (version 4.12), but the electron temperatures are close to the earlier values.  A similar equation accommodates the effects of the electron density \citep[Equation 35,][]{2013ApJS..207...21N}.

Similarly, from \citetalias{2014ApJ...786..155N}, the abundance of O$^+$ can be expressed in terms of the observed fluxes from the [O~{\sc ii}] $\lambda$3726 and $\lambda$3729 lines,
\begin{equation}\label{eq4}
\frac{n_{O^+}}{n_{H^+}}  = \frac{\text{flux}(\text{O{\sc ii})}}{\text{flux}(\text{H}_\beta)}.g_{1(\text{OII})}.\sqrt{T_e}.\alpha^{eff}_B (\text{H}_{\beta}).\text{exp}(E_{12(\text{OII})}/(kT_e)) \times115885.4/(E_{12(\text{OII})}.\Upsilon_{12(\text{OII})}) ,
\end{equation}
where, in this case, $T_e$ is the apparent electron temperature derived from the [O~{\sc ii}] ratio \citep[see][]{2013ApJS..207...21N} using the  ratio of the 7320-30 \AA~ lines to the 3726/9 \AA~lines.  When the NIR lines are not available, it is possible to derive an expression for the [O~{\sc ii}] electron temperature from the Mappings photoionization models as a polynomial in terms of \emph{total} oxygen gas-phase abundance,
\begin{equation}\label{eq5}
T_e(\text{O{\sc ii}})=T_e(\text{O{\sc iii}}) \times (3.0794 - 0.086924~\zeta - 0.1053~\zeta^2 + 0.010225~\zeta^3)
\end{equation}
where $\zeta$=12+log(O/H) \footnote{Here we use $\zeta$ for the oxygen numerical abundance measured from spectra, to distinguish it from Z, the oxygen abundance by mass}. This equation is used iteratively, as discussed in \citetalias{2014ApJ...786..155N}, starting by using the O$^{++}$ abundance as the total oxygen abundance. This process converges rapidly in 5 iterations or less. \cite{1992AJ....103.1330G} and  \cite{2012MNRAS.426.2630L} have used a simpler approach, expressing the low ionization zone temperature (effectively the [O~{\sc ii}] temperature) in terms of the [O~{\sc iii}] temperature, which does not require iteration. 

Equation \ref{eq5a} shows the expression used by \cite{2012MNRAS.426.2630L}:
\begin{equation}\label{eq5a}
T_e(\text{O{\sc ii}})=T_e(\text{O{\sc iii}}) +450 -70\times \exp\left[(T_e(\text{O{\sc iii}})/5000)^{1.22}\right]
\end{equation}
Equation \ref{eq5a} gives total oxygen abundance values close to those from iterating Equation \ref{eq5}. Values determined for  oxygen abundances are not exact, because of the nature of the approximations used, the calculated values for oxygen abundances depend on the photoionization models used to build the models, and the use of a model derived from a single value of the ionization parameter, $q$. We tested the two methods (Equations \ref{eq5} and \ref{eq5a}) against artificial data from Mappings, and find that they generate total oxygen abundances within 1\% of the input values. The iterative approach (Equation \ref{eq5}) is marginally the more consistent of the two over a range of ionization parameter values.   Using the polynomial and exponential fits to the parameters in these equations from \citetalias{2014ApJ...786..155N} (Equations 6-9) , we can calculate the total gas-phase oxygen abundance from the observed NUV [O~{\sc ii}] and optical [O~{\sc iii}] lines.

\subsection{Limitations due to estimates of the  [O~{\sc ii}] temperatures and abundances}
There are at least four sources of uncertainty in  determining [O~{\sc ii}] electron temperatures, and in general these are significantly less accurately measured than the [O~{\sc iii}] electron temperatures.  First, directly estimating the electron temperature requires measurements of the 3726,9\AA~ and 7320,30\AA~ lines in the ultraviolet and infrared, which are not always available in any given set of observations---the SDSS, for example, does not measure the UV lines unless they are red-shifted into the spectroscopic passband.  Second, both sets of lines are usually subject to signal-to-noise problems.  Third, both Equations \ref{eq5} and \ref{eq5a} depend on the model parameters used. The models themselves depend on the computed collision strengths.  In the case of  O~{\sc ii}, there are several collision strength data sets \citep{1998JPhB...31.4317M, 2006MNRAS.366L...6P, 2007ApJS..171..331T, 2009MNRAS.397..903K} which are not in full agreement, resulting in slightly different calculated electron temperatures. These propagate into the estimation of  O~{\sc ii} abundance values, and thus into the total oxygen abundance. Finally, the electron temperature formula determined from single slab models varies with the ionization parameter, $\log(q)$, which differs for each observed object. The observational problems for [O~{\sc ii}] temperatures were also discussed by \cite{2003ApJ...591..801K}.

In this work we have used the following sources: in Mappings IV we use the Tayal 2007 collision strengths; the generic [O~{\sc ii}] electron temperature formula derived from single slab Mappings IV models used in \cite{2013ApJS..208...10D}; and the iterative method to calculate total oxygen abundance (Equation \ref{eq5}). Uncertainties arising from the flux measurements remain the largest source of error.

\subsection{Results from published data}
Table \ref{t1} shows the data from \citetalias{2014ApJ...786..155N} for 16 objects from the SIGRID sample for which oxygen abundances are available from  both the direct T$_e$ method (Equations \ref{eq1}, \ref{eq4} above) and the strong line methods using the diagnostic grids from \cite{2013ApJS..208...10D}. We have used the diagnostic grids for  log(N{\sc ii}/S{\sc ii}) versus log(O{\sc iii}/S{\sc ii}) and log(N{\sc ii}/S{\sc ii}) versus log(O{\sc iii}/H$\beta$) as they provide the most consistent results, as noted in \citetalias{2014ApJ...786..155N}.  Column 5 in Table \ref{t1} shows the difference between the direct and strong line methods. In general the strong line methods give higher oxygen abundances, with a mean excess of 0.149 dex.  This is consistent with the findings of \cite{2012MNRAS.426.2630L}. The primary source of this difference is that the diagnostic grids we use here include the total oxygen content, that is, the gas-phase oxygen plus the oxygen incorporated in dust grains, whereas the direct method yields only the gas-phase oxygen.  This contributes a difference of 0.07 dex, or half the observed amount.  As we will show below, there are other sources contributing to the observed differences.

\ctable[
caption= {Apparent electron temperatures, derived gas-phase oxygen abundances, strong line abundances and abundance differences for SIGRID objects, from \citetalias{2014ApJ...786..155N}},
doinside=\small,
mincapwidth=0.8\textwidth,
pos=h,
captionskip=3pt,
label=t1,
]{lllll}
{
\tnote[1]{T$_e$ and 12+log(O/H) values from \citetalias{2014ApJ...786..155N}, Table 3}
\tnote[2]{Strong line values are the average of the new grids involving the log(N~{\sc ii}/S~{\sc ii}) and log(N~{\sc ii}/O~{\sc ii}) ratios, where available, from \citetalias{2014ApJ...786..155N}, Table 4}
\tnote[3]{delta is the difference between columns 3 and 4.}
} 
{\toprule[1.5pt]
Object			&	T$_e$ (K)\tmark[1]		&	Direct 	&	 Strong line\tmark[2]	&	delta\tmark[3]	\\
\midrule
J0005-28			&	14719	&	7.847	&	 8.012	&	 0.165 	\\
J1152-02A		&	12248	&	8.151	&	 8.088	&	-0.063 	\\
J1152-02B		&	12722	&	8.094	&	 7.981	&	-0.113 	\\
J1225-06s2		&	16560	&	7.499	&	 7.929	&	 0.430 	\\
J1328+02			&	14846	&	7.867	&	 8.132	&	 0.265 	\\
J1403-27			&	14021	&	7.939	&	 8.013	&	 0.074 	\\
J1609-04(2)		&	10431	&	8.345	&	 8.055	&	-0.290 	\\
J1609-04(5)		&	14234	&	7.959	&	 8.137	&	 0.178 	\\
J2039-63A		&	14383	&	7.966	&	 8.087	&	 0.121 	\\
J2039-63B		&	13982	&	7.894	&	 8.145	&	 0.251 	\\
J2234-04B		&	14029	&	7.897	&	 8.110	&	 0.213  	\\
J2242-06			&	13626	&	7.921	&	 7.983	&	 0.062 	\\
J2254-26			&	12870	&	8.090	&	 --- 		&	 --- 		\\
J2311-42A		&	12582	&	8.091	&	 8.176	&	 0.085 	\\
J2311-42B		&	13451	&	8.011	&	 8.122	&	 0.111	\\
J2349-22			&	13835	&	7.858	&	 7.925	&	 0.067	\\
\addlinespace[5pt]\bottomrule[1.5pt]\addlinespace[5pt]}

\FloatBarrier

\section{The [O~{\sc iii}] electron temperature--abundance relation}

Figure \ref{teza} is a plot of of the theoretical [O~{\sc iii}] electron temperature versus oxygen abundance for Mappings isobaric model curves for $\log(q)$=8.50,  $\kappa=\infty$, for optically thick nebulae, and an electron density of $\sim$7.5 cm$^{-3}$ ($\log(P/k)$=5). The figure shows the observational points derived from the SIGRID survey (yellow circles), selected SDSS DR3 data from \cite{2006A&A...448..955I} (black circles),  \cite{2012A&A...546A.122I} (small beige circles), and \citet{2009A&A...508..615L} (blue circles). Mean error bars for each sample are shown. Detailed error bars have been omitted for clarity, but are shown in \citetalias{2014ApJ...786..155N} and \cite{2006A&A...448..955I}. The effects of  hydrogen optical depth, excitation sources, pressure, ionization parameter, dust, and $\kappa$ non-equilibrium electron distributions are discussed in detail in Sections 4 through 9 (below). While the uncertainties in the plotted values are larger than the detailed trends in the data, nonetheless, the data points provide a useful basis for examining the impact of different physical parameters on the behavior of an \HII region.

The 125 objects selected from the SDSS DR3 data in \cite{2006A&A...448..955I} exclude those with large uncertainties and those without flux data for the [O~{\sc ii}] 3726 and 3729\AA~ lines, and oxygen abundances in the range for which the model curves apply: 7.32~$<$ 12+log(O/H) $<$~8.69. The data selected from \cite{2012A&A...546A.122I} are those with error bars less that 15\% of the measurements, and oxygen abundances in the range for which the model curves apply. The total gas-phase metallicities were calculated from the published data using the formulae in Equations \ref{eq1} and \ref{eq4}. We found that the spread of values using these equations is less than the spread using the original formulae in  \cite{2006A&A...448..955I}, suggesting the new method may give more reliable results, or that using the observed [O~{\sc ii}] 7320/30 \AA~ fluxes is prone to higher uncertainty, as suggested by \cite{2003ApJ...591..801K}. Note that the horizontal axis is the gas-phase rather that the total (gas+dust) abundance, which differ by 0.07 dex.

\begin{figure}[htbp]
\centering
\includegraphics[width=0.8\hsize]{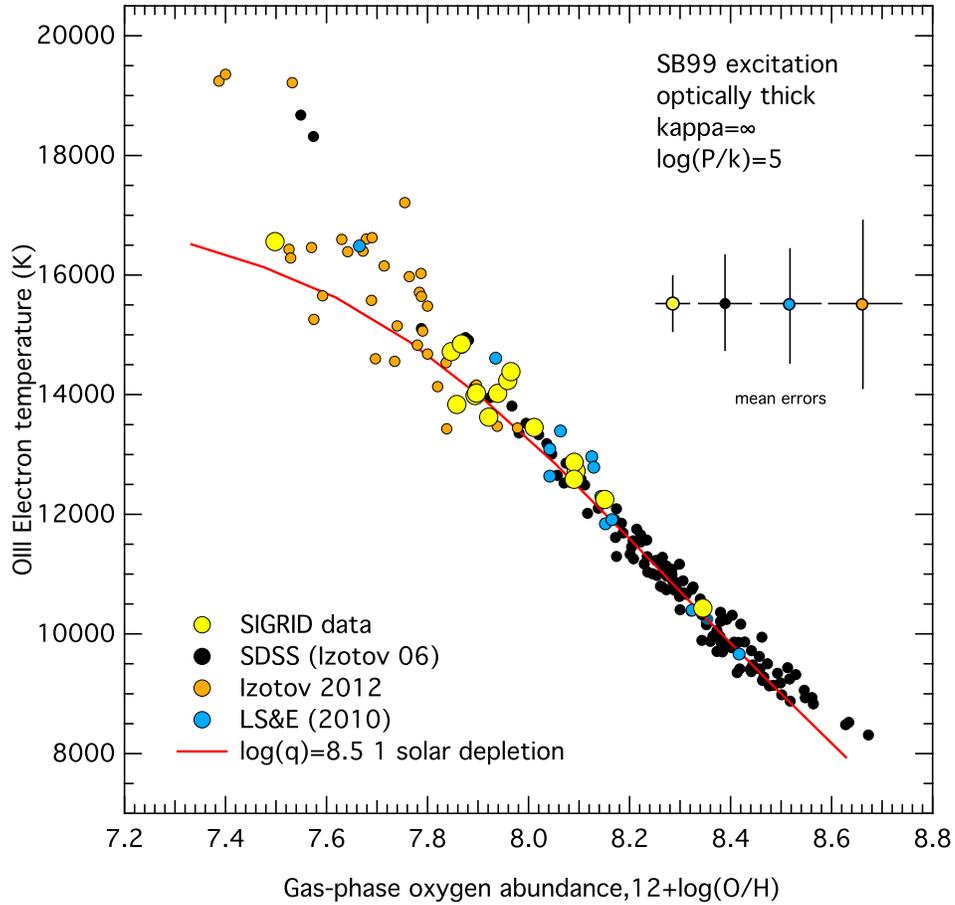}
\caption{Observed apparent electron temperature versus gas-phase oxygen abundances from the SIGRID data (\citetalias{2014ApJ...786..155N}), \cite{2006A&A...448..955I,2012A&A...546A.122I} and \cite{2009A&A...508..615L}, and Mappings photoionization model curve (red) for $\log(q)$=8.50 (see text).}\label{teza}
\end{figure}

The data from \cite{2006A&A...448..955I,2012A&A...546A.122I} were chosen because they are consistent sets of data from single observational sources. The sparse low metallically coverage of the SDSS data is augmented using data from \cite{2012A&A...546A.122I}. The two sets cover the range from 1/50 solar to 1 solar.  This spans the range covered by the models used, 1/20 solar  to 1 solar. (The \cite{2006A&A...448..955I} data extends above 1 solar metallicity, but we have truncated the selection for comparing with the models.)  Additional data from \cite{2009A&A...508..615L,2010A&A...517A..85L} were also investigated. The results are nearly identical. Using their published line fluxes, there were 13 objects for which data was available to use the same methods to calculate T$_e$ and 12+log(O/H) that we use here for the SIGRID and Izotov data.  The results of these calculations exactly overlie the other data. The mean error bars shown were calculated from the flux errors in the original observations, propagated through the temperature and abundance calculations, and are somewhat larger than those reported in the source papers.

\FloatBarrier

Figure \ref{teza} shows a remarkable fit of the Mappings model to the observed data (notwithstanding the size of the uncertainties), although the data points with lower oxygen abundances ($<$ 8.0) tend to lie increasingly above the model curve with lower abundance. An analytical fit to the model curve is given in Equation \ref{eq7}, which applies for oxygen abundances in the range 7.4~$< 12+\log(O/H) <$~8.7. This equation provides a quick means of estimating abundance once the [O~{\sc iii}] electron temperature has been measured.

\begin{equation}\label{eq7}
\zeta = 1.0324 + 30.364\times t_4 - 43.019\times t_4~^2 + 25.694\times t_4~^3 - 5.6791\times t_4~^4
\end{equation}
where $\zeta$= 12+log(O/H), $t_4$=T$_e$/10000, with an average fit error in $\zeta$ of 0.004 dex. Note that this fit should not be used for extrapolation outside the model limits.

Figure \ref{teza} shows two SDSS data points at low abundance and high temperature from the \cite{2006A&A...448..955I} data.  Both points refer to a single object (observed twice in SDSS), HS0837+4717.  This is an extraordinary object, with very high nitrogen abundance, and the presence of over 100 WN stars in the central cluster \citep{2004A&A...419..469P, 2011A&A...532A.141P}. The more recent IFU spectral analysis by P\'erez-Montero gives a significantly lower apparent electron temperature than the SDSS observations, although it is still high (see Figure \ref{tezb}).

Figure \ref{teza} is key to exploring the factors influencing the behavior of electron temperature with oxygen abundance, using the data from \cite{2006A&A...448..955I,2012A&A...546A.122I}, \citet{2009A&A...508..615L}, and the SIGRID data from \citetalias{2014ApJ...786..155N} as guides to the actual behavior of \HII regions. In the figure, it is clear that the model suggests the electron temperature falls below a linear trend as the oxygen abundance decreases.  This can be understood by considering the physical processes occurring in spherically symmetric ``Str\"omgren sphere'' \HII regions. In the (innermost) region of highest excitation, the temperature is highest, but the volume is relatively small compared to cooler regions further from the excitation source.  The observed fluxes are the volume weighted averages.  The larger volumes of cooler lower ionization (outer) regions dilute the effect of the hotter inner regions.  The net result is that for any particular excitation source, there is a limit to the observed apparent electron temperature that can be reached in a spherical model, no matter how low the oxygen abundance, and this is lower than the temperature of the innermost ionized region.  As a result, the apparent  electron temperature would be expected to fall below a simple linear trend as oxygen abundance decreases, which is what our models show.  The only variables are the slope of the trend at higher metallicities, and the maximum apparent electron temperature that can be reached. These geometric arguments apply to any \HII region, although the detailed averaging processes in complex nebulae can only be approximated using simple spherical models. This model also ignores other physical parameters that can increase the apparent electron temperature. It should be noted that in real \HII regions or in models fitting such real regions, the ``central'' regions may well not be the hottest \citep[See, for example][]{1978A&A....66..257S,2013A&A...551A..82S,2009ApJ...700..654E}. However, for the purposes of this study which explores the effects of individual physical parameters, we use models that are by definition hottest at the centre.

This investigation was stimulated by the apparent inability of simple models to match the observed T$_e$---abundance data. The problem has been known for over a decade in the context of low metllicity blue compact dwarf galaxies. It was explored by \cite{1999A&A...351...72S} in the context of the complex nebulosity in I Zw18. However, the mismatch is not confined to that object, but is generic to most low metallicity \HII regions. More recently, \cite{2008A&A...478..371P} modeled the complex structures in I Zw 18 and concluded that  simpler models did not adequately account for small-scale gas density fluctuations, which can be important when the collisional excitation of hydrogen contributes significantly to the cooling of the gas. In this work we have taken a different approach, in part because the structures in the \HII regions we consider are not resolved. The Mappings code takes detailed account of all the excitation, ionization, heating, and cooling mechanisms and the effects of dust, present in \HII regions, and their complex interactions, rather than attempting to taylor a fit to indvidual objects. In this paper we use Mappings with a Str\"omgren Sphere model to explore separately the effects of several individual parameters.  These include the optical depth of hydrogen in the nebula, the pressure/density, the ionization parameter, the nature of the ionizing radiation from the central star cluster, the effect of dust, and non-equilibrium $\kappa$ electron energy distributions.  All of these affect the shape of the model curves for apparent electron temperature versus oxygen abundance.  We explore these effects and how well they fit the observations in the following sections.

\section{The effect of optical depth at the Lyman limit}
The first important parameter is the optical thickness (at the Lyman limit in hydrogen at 912\AA) in the \HII regions. In optically thin nebulae we find higher values for T$_e$ at low metallicities than predicted for optically thick  regions.  We also identify parameters that might indicate optically thin \HII regions. The question has been addressed by \cite{1979A&AS...35..111K} in the context of clumpy nebulae with optically thin condensations, and more recently by \cite{2012ApJ...755...40P}, who measured the optical depth of \HII regions in the Magellanic Clouds.

With the present data, individual \HII regions are not resolved as well as the nearby regions in the Magellanic Clouds, so a more general approach it necessary.  Figure \ref{opthin} shows three narrow-band images of the Rosette nebula (NGC 2237).  The central star cluster and stellar wind evacuated central region are clearly visible in the [O~{\sc iii}] and H$\alpha$ images. There is some suggestion that the area in the top left quadrant is optically thin, which lacks strong [S{\sc ii}] flux.  An observer looking at this object from a distance sees some of the optical spectral lines arising from regions within the ionized region (optically thin), and some from regions where all the ionizing photons from the central star cluster have been absorbed (optically thick).  In our model nebulae, the brightness of the inner regions is significantly less than from the outer regions, so that the net effect is to average over a range of optical depths, but skewed in favor of the brighter, optically thick regions.
\begin{figure}[htbp]
\centering
\includegraphics[width=1.0\hsize]{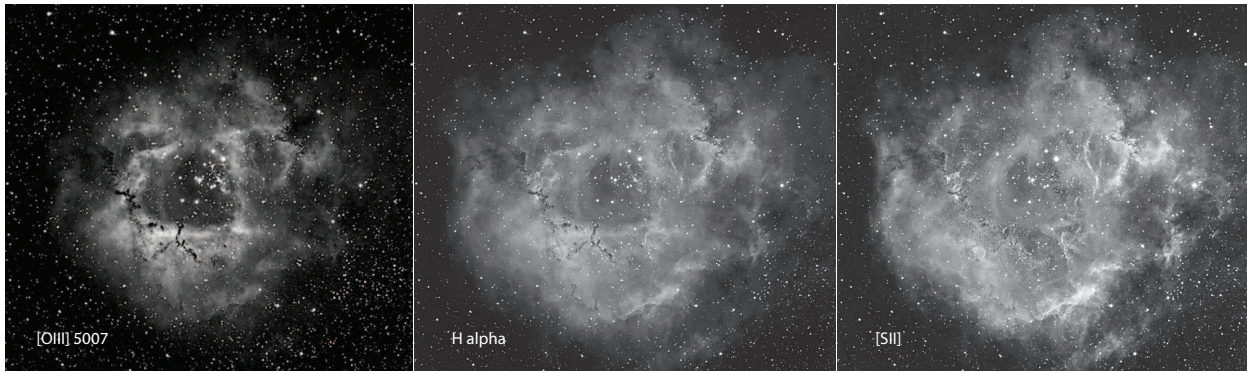}
\caption{Rosette Nebula (NGC 2237) \HII region, in [O~{\sc iii}] 5007\AA, H$\alpha$, [S~{\sc ii}] 6716/31 \AA~narrow band images.  The [O~{\sc iii}] is predominantly in the inner zone and the [S~{\sc ii}] in the outer zone.  The central star cluster and stellar-wind-evacuated core are apparent in the [O~{\sc iii}] and H$\alpha$ frames. There is also evidence of an optically thin region in the upper left quadrant, where the [S~{\sc ii}]  is not prominent. Image reproduced with permission from Steven Coates (\protect\url{http://coatesastrophotography.com})}\label{opthin}
\end{figure}
\FloatBarrier

To illustrate the effect of an optically thin nebula, we use the Mappings code to generate the relation between observed apparent electron temperature and oxygen abundance, taking an extreme case where the optical depth, $\tau$=1.  Figure \ref{tezb} shows the same data as in Figure \ref{teza}, with the addition of the $\tau$=1 model curve (dashed black), and the data from \cite{2011A&A...532A.141P} for the extreme object (HS0837+4717). It is clear that this optically thin model ($\tau$=1 and $\log(P/k)$=5) is capable of explaining some (but not all) of the observed higher values of T$_e$ in the observations.

\begin{figure}[htbp]
\centering
\includegraphics[width=0.8\hsize]{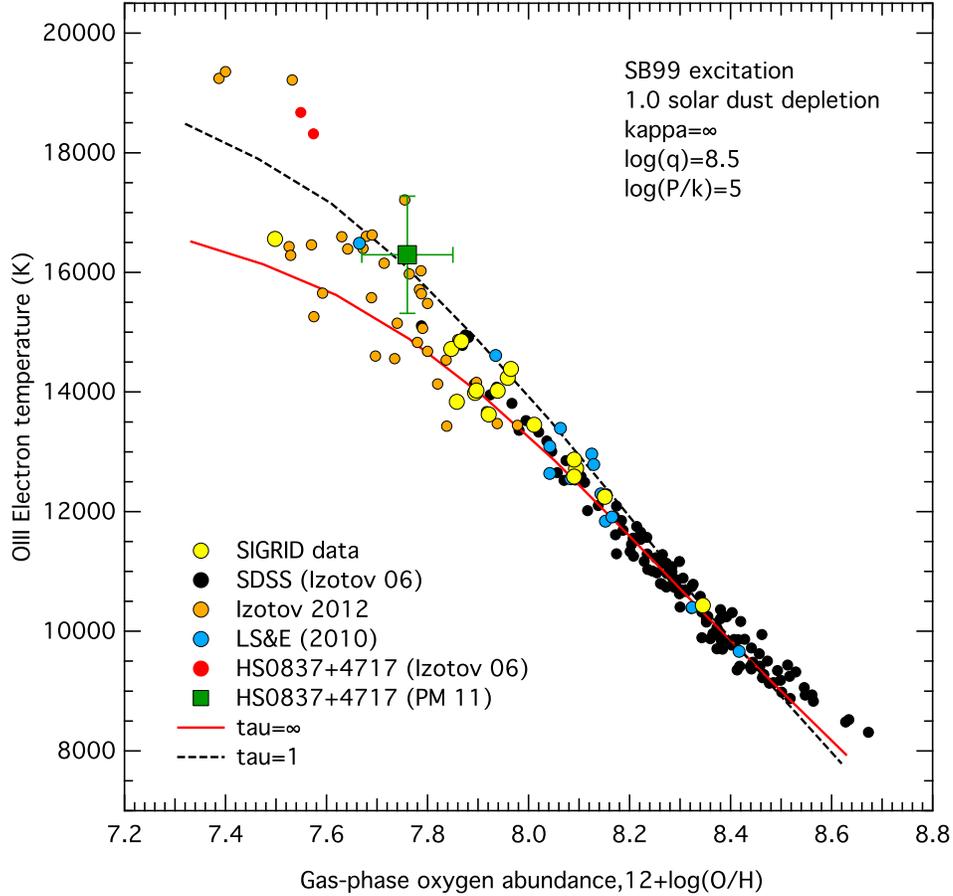}
\caption{Observed apparent electron temperature versus oxygen abundance, with Mappings model curves for the optically thick and $\tau$=1 cases (see text). The red SDSS data points for the blue compact dwarf HS0837+4717 correspond to the more precisely measured data from \cite{2011A&A...532A.141P} (green square).}\label{tezb}
\end{figure}
\FloatBarrier

\subsection{Nebular structure}

Figure \ref{fnorm} shows the normalized fluxes of different ions as a function of radius, calculated using the Mappings IV.1.2 photoionization code, with similar settings to Figure \ref{fraco}. The 4363\AA~line arising from the higher energy $^1$S$_0$ O~{\sc iii} level is enhanced compared to the 5007\AA~ line in the inner regions of the nebula where the electron temperature is higher, as shown by the  T$_e$ curve (black line). Similar diagrams have been published by other authors, for example, Figure 1 in \cite{2012ApJ...755...40P}

\begin{figure}[htbp]
\centering
\includegraphics[width=0.55\hsize]{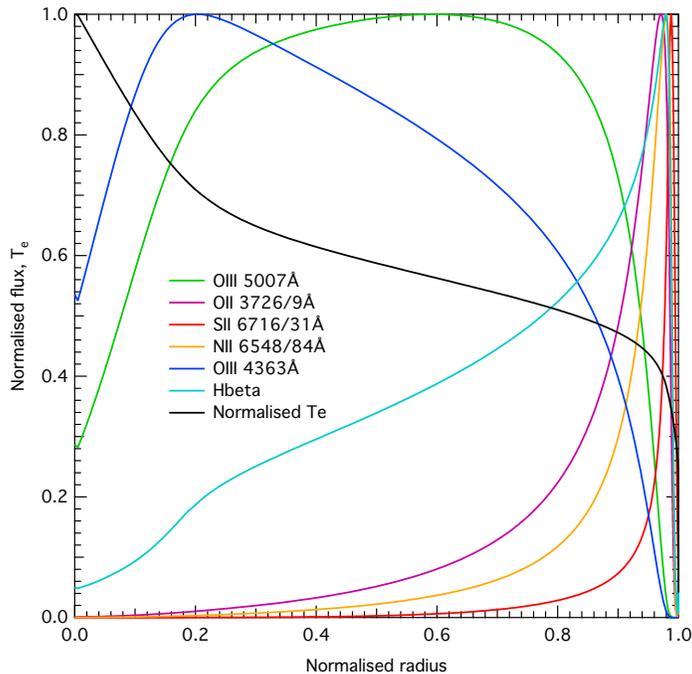}
\caption{Normalized fluxes and T$_e$, versus radius, low metallicity (0.1 solar)}\label{fnorm}
\end{figure}

It is worth noting the the curve in Figure \ref{tezb} for low optical depth (dashed black) falls below the optically thick curve (red) with increasing metallicity. The physical origin of this cross-over is the nature of the thermal balance inside an \HII region at different metallicities.  At low metallicity, as shown by the black curve in Figure \ref{fnorm}, the temperature decreases monotonically with increasing radius. Hydrogen and helium play the main role in absorbing high energy photons from the central star cluster. They are more efficient at absorbing the softer photons, so the spectral energy distribution hardens with increasing radius, and some of the hard photons may leak out of the nebula.  At higher metallicities, the heavier elements start to play an important role in the heating and cooling in the outer regions of the nebula. Due to their numerous energy levels, they are able to absorb the higher energy photons in the harder SED more efficiently than hydrogen and helium. This causes additional heating in the outer regions of the nebula, leading to a rise in temperature, as the heating and cooling processes balance and establish an equilibrium. Consequently, as the metallicity increases, the temperature curve shown in Figure \ref{fnorm} starts to flatten out, and can actually increase in the outer regions of the nebula at higher metallicities. For optically thin nebulae, the outer regions of the nebula are truncated, so the hotter outer zones are removed, and the apparent net temperature falls, leading to a greater decrease of apparent temperature at higher metallicity with decreasing optical depth, compared to the behavior at lower metallicity. Consequently the low optical depth curve falls away faster with increasing metallicity than the optically thick curve, as shown in Figure \ref{tezb}.

\FloatBarrier

\subsection{Optical depth diagnostics}

Figure \ref{fnorm} shows how the majority of the [N~{\sc ii}], [S~{\sc ii}] and---to a slightly lesser extent---[O~{\sc ii}] emissions are confined to the outer regions of the nebula.  If the nebula is optically thin, the outer regions are truncated, and the total emissions from these outer species are substantially reduced, whereas the [O~{\sc iii}] lines and H$\beta$  are emitted throughout the body of the nebula and are only gradually reduced by truncating the outer layers.  This behavior---the lack of bright [N~{\sc ii}] and [S~{\sc ii}] edges to \HII regions---was used by \cite{2012ApJ...755...40P} in analyzing the optical thickness of \HII regions in the Magellanic Clouds.  This suggests that spectral line ratios might be useful diagnostics of low optical depth in nebulae where the edges of \HII regions are not well resolved. Candidate line ratios would be [O{\sc ii}]/[O{\sc iii}], [N{\sc ii}]/[O{\sc iii}] and [S{\sc ii}]/[O{\sc iii}].

Of these three opacity diagnostic options, the [S{\sc ii}]/[O{\sc iii}] ratio appears to be the most useful, for several reasons. The [S~{\sc ii}] lines are well resolved from the strong H$\alpha$ Balmer line. They are not subject to the nitrogen enrichment processes that cause an increasing spread in [N~{\sc ii}] fluxes at low oxygen abundance. The [S~{\sc ii}] lines are almost invariably present in the spectra of \HII regions at good signal-to-noise ratios, and the [S~{\sc ii}] emission region is more strongly concentrated at the lower excitation outer edges of an \HII region than the [O~{\sc ii}]. To illustrate this diagnostic, Figure \ref{s2o3} shows log([S{\sc ii}]/[O{\sc iii}$_{4959}$]) versus oxygen abundance, for Mappings models of several values of optical depth and $\log(q)$, and the observed values of the SDSS objects from \cite{2006A&A...448..955I}, the emission line galaxies from \citet{2012A&A...546A.122I} and the SIGRID objects from \citetalias{2014ApJ...786..155N}\footnote{We use the [O~{\sc iii}$_{4959}$] line in these calculations as the brighter  [O~{\sc iii}$_{5007}$] line fluxes are not reported in \cite{2006A&A...448..955I}.}.

\begin{figure}[htbp]
\centering
\includegraphics[width=0.8\hsize]{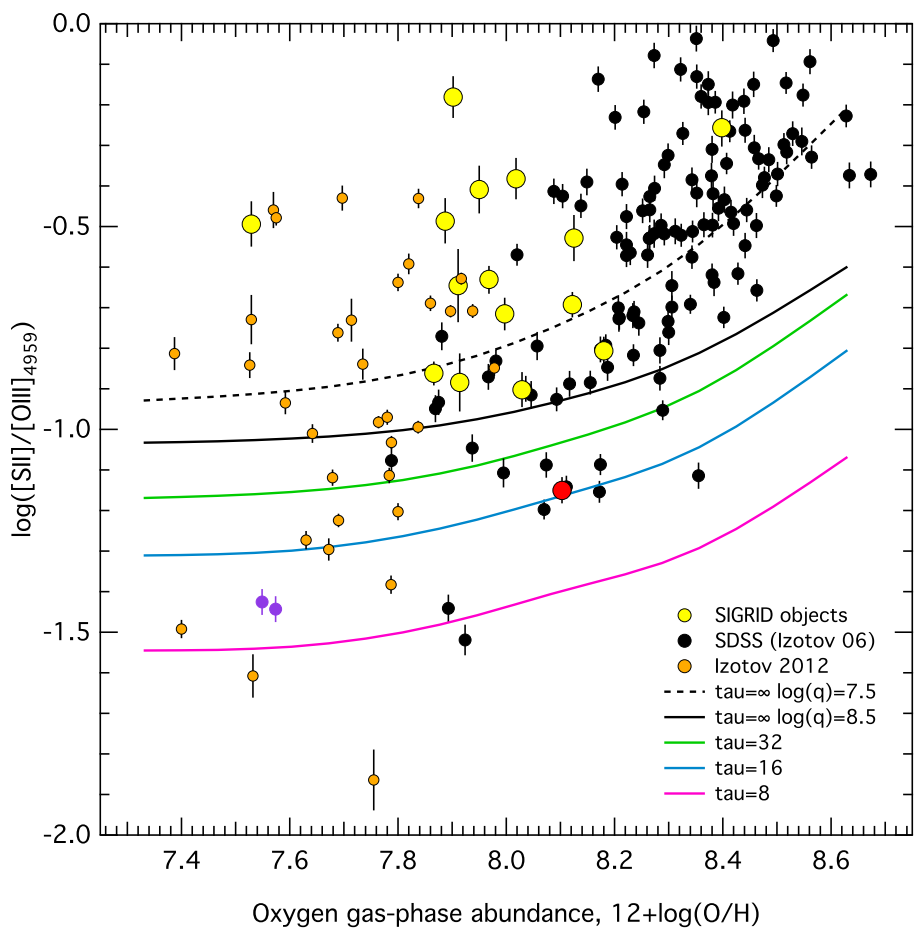}
\caption{The log([S{\sc ii}]/[O{\sc iii}$_{4959}$]) diagnostic for optical depth, comparing observations with model curves. Objects below the black solid line ($\log(q)$=8.5) are optically thin candidates, and those below the dashed black line ($\log(q)$=7.5) are probable candidates. The purple circles are the two SDSS observations of HS0837+4717 and the large red circle is the SIGRID object J2254-26. }\label{s2o3}
\end{figure}

It is evident that the majority of the observations fall above the optically thick line for $\log(q)$=8.5, and roughly half fall above the same curve for $\log(q)$=7.5, illustrating the interaction of the line ratio interpretations with the ionization parameter. The error bars are all smaller than the separation of the model curves.  Thus it is not possible to distinguish between the effect of the ionization parameter, $\log(q)$, and the optical thickness for borderline objects.  However, it is also clear that more extreme objects can be identified, such as J2254-26 from the SIGRID list (red circle) and the SDSS double points for HS0837+4717 (purple circles) (\cite{2012ApJ...755...40P} did not obtain fluxes for the S~{\sc ii} lines due to its redshift). The scarcity of objects with optical depth $\tau <$ 8 may be due to the flux weighting of the contributions from parts of the nebular with different optical depths, which is dominated by the brighter, more optically thick regions.

As a comparison, Figure \ref{o2o3} shows the same curves for log([O{\sc ii}$_{3727}$]/[O{\sc iii}$_{4959}$]).  The theoretical curves for higher values of $\tau$ become degenerate, and while they show the same general trend,  the diagram is not in complete agreement with the log([S{\sc ii}]/[O{\sc iii}$_{4959}$]) and implies lesser optical depth. Given the weighting from the brighter, thicker regions of the model nebula, it is likely that the log([O{\sc ii}]/[O{\sc iii}]) diagnostic underestimates the optical depth. The twin SDSS points for HS0837+4717 (purple circles), the datum for this object from \cite{2012ApJ...755...40P}, and SIGRID object J2254-26 (red circle) again stand out as optically thin. It is also worth noting that the model curves are derived from a simple Str\"omgren Sphere model, whereas the observed data arise from more complex structures.

\begin{figure}[htbp]
\centering
\includegraphics[width=0.8\hsize]{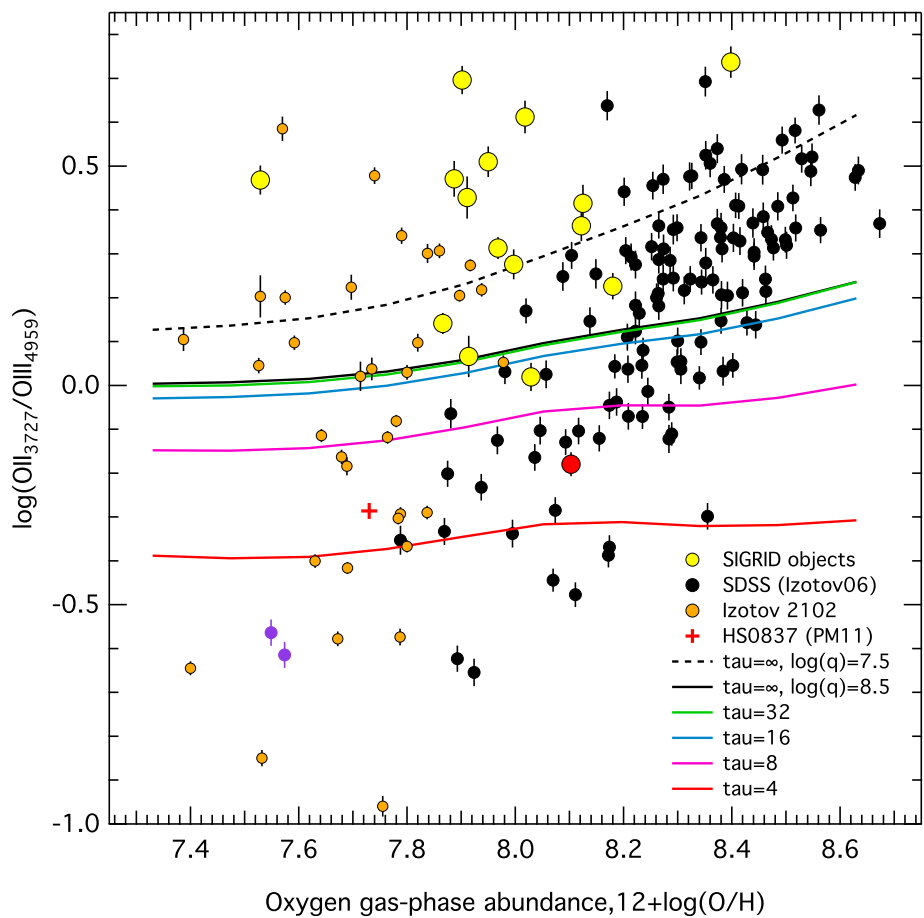}
\caption{The log([O{\sc ii}]/[O{\sc iii}]) diagnostic for optical depth, comparing observations with model curves from Mappings. Objects below the black solid line are optically thin candidates. The red circle is SIGRID object J2254-26 and the purple circles are the SDSS observations for HS0837+4717.}\label{o2o3}
\end{figure}

\FloatBarrier

\subsection{Combining spectra from different optical depths}

In real \HII regions, the emitted spectra are likely to arise from regions with different optical depths.  For distant objects where the individual regions are not resolved, the spectra are hybrid, combining spectra from a range of optical depths.  Figure \ref{taurat} shows the behavior of the [O~{\sc iii}] apparent electron temperatures from these hybrid sources.  The solid curves correspond to mixtures of spectra from optically thick regions with those from optically thin regions where $\tau$=1, in ratios between 1:8 and 8:1.  The components of the composite spectra are weighted according to the H$\beta$ luminosity of each spectrum. In this diagram, the electron temperatures computed from the combined spectra are normalized to the temperature curve for optically thick regions.  All models are for $\kappa=\infty$ and $\log(q)$=8.5. Single optically thin regions with $\tau$=1, 2, and 4 are shown (dashed lines) for comparison.  The hybrid spectra give temperature/abundance curves that do not exactly match any single value of $\tau$, but below an oxygen abundance of $\sim$8.0, the curve for $\tau$=2 corresponds quite closely to a mixture of 3.5:1 (thin to thick), and the curve for $\tau$=4 to a mixture of 1.5:1 (thin to thick). The cross-over point of the hybrid curves corresponds to that in Figure \ref{tezb} and is the result of the changing fluxes with increasing metallicity as generated by the models, and the normalization of the averaged T$_e$ to the optically thick model.

\begin{figure}[htbp]
\centering
\includegraphics[width=0.8\hsize]{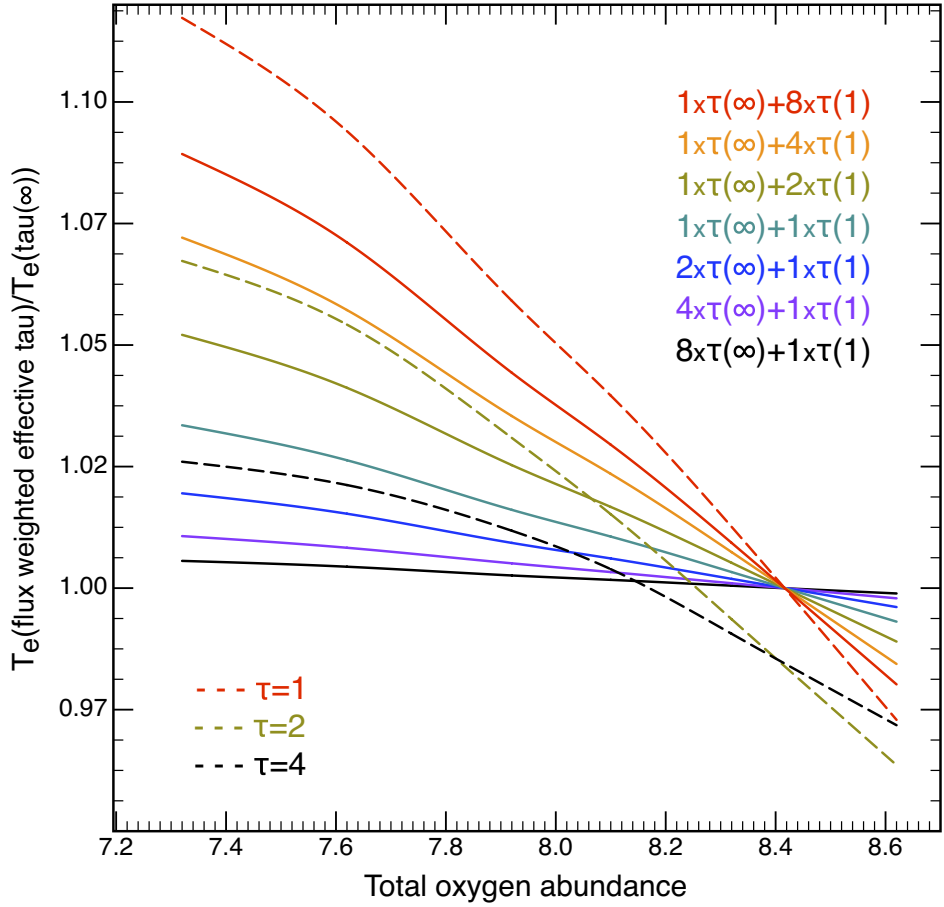}
\caption{The effect of combining spectra from different optical depths. The cross-over point of the hybrid curves corresponds to that in Figure \ref{tezb} and is the result of the changing fluxes with increasing metallicity as generated by the models, and the normalized T$_e$. See Section 4.1 for a detailed discussion.}\label{taurat}
\end{figure}

\FloatBarrier

\section{Effect of the UV Spectral Energy Distribution}

The Mappings models used here are calculated for an oxygen abundance range from 0.05 to 1.0 solar, based on the Starburst99 code from \cite{1999ApJS..123....3L}, with a Salpeter initial mass function, $dN/dm \propto m^{-2.35}$, a lower mass cut-off at 0.1 $M_\odot$ and an upper mass cut-off at 120 $M_\odot$, with continuous star formation \citep[described in][]{2000ApJ...542..224D, 2013ApJS..208...10D}. We use a stellar population of 10$^6$ M$_\odot$ to avoid stochastic effects on the shape of the EUV continuum. The Starburst99 code does not provide good coverage for very low metallicities and hot star clusters, so we have used a blackbodies at 50,000K, 75,000K and 100,000K as high temperature sources\footnote{For this work we used version 6 of Starburst99. Version, 7, just released, does include important new atmospheres \citep{2014arXiv1403.5444L,2014arXiv1402.0824L}, but not yet at the metallicities we need to take this work to lower abundances.}.  Figure \ref{tezbb} shows the Mappings model results for these sources, compared to the Starburst99 model source.  It is evident that while high temperature excitation sources can generate higher apparent electron temperatures at low oxygen abundances, the overall fit to the observed data for blackbodies is not as good as the Starburst 99 model. However, for any particular object, such as HS0837+4717, a higher excitation source may well contribute to a higher apparent electron temperature. We have also explored the effect of single hot star models derived using the WMBasic code \citep[][and related papers]{2001A&A...375..161P}, with results very similar to the blackbody models.

A problem we face in exploring the effects of the central ionization source is the lack of spectral energy distribution models at low metallicities.  There are two important deficiencies.  The first is the lack of evolutionary tracks for a range of massive low metallicity stars.  As a result, our models are limited to oxygen abundances at and above 0.05 solar.  The second matter that needs attention, once we have adequate low-metallicity stellar models, is to build small-number statistical models for small star clusters likely to be found in small \HII regions. The spread of nebular abundances at low metallicities \citep[Figure 6,][] {2004ApJ...613..898T} for a given stellar mass, and a similar spread in log(N/O) values \citepalias[see][]{2014ApJ...786..155N} for a given oxygen abundance, which exceed the uncertainties in many cases, suggest that stochastic effects are important in modeling stellar masses in small \HII regions in small galaxies.

Even with appropriate low metallicity stellar models, however, the energy distribution from the  cluster exciting the \HII region may not completely account for the excess [O~{\sc iii}] 4363\AA~ fluxes observed in low metallicity nebulae.  For example, \cite{1999A&A...351...72S} found that, even taking into account extreme ionizing radiation sources, photoionization models could not account for the 4363\AA~ flux in I Zw 18. Our models are consistent with this conclusion.

\begin{figure}[htbp]
\centering
\includegraphics[width=0.8\hsize]{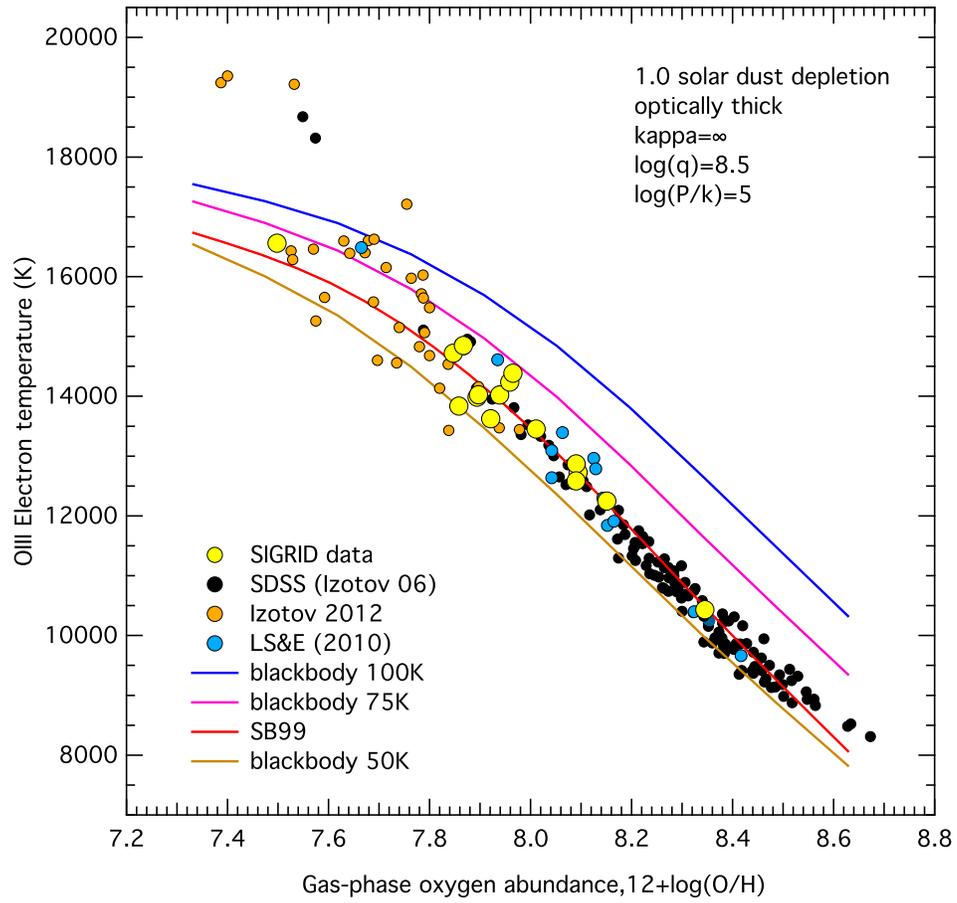}
\caption{Mappings models using different excitation sources.}\label{tezbb}
\end{figure}

\FloatBarrier

\section{Pressure effects}

The electron density, n$_e$, in an \HII region can have important effects on the emitted spectrum. The parameter varies throughout the nebula, but it is useful to explore how setting a global value for the n$_e\times$T$_e$ affects the electron temperature measurements.  In our Mappings simulations, we use  an isobaric model (constant pressure) using the parameter P/k  (where P is the pressure and k is the Boltzmann constant).  P/k = n$_{tot}$$\times$T$_e$, where n$_{tot}$ is the total particle number density, and n$_{tot}$ $\approx$ n$_e$ $\times$ 2.07 at low metallicities. Thus, for isobaric models, the product of the total number density and the electron temperature is constant. While the electron densities \textit{per se}~are not constant in these models, the $\log(P/k)$ values, together with the electron temperatures, set the electron densities: e.g., if the electron temperature is 10,000K and $\log(P/k)$=6, the electron density is $\sim$48.4 cm$^{-3}$. 

Fixed pressure models are more physically realistic than fixed density (isochoric) models, in part because they allow the density to increase near the ionization front. At typical temperatures and densities, the sound crossing time for an \HII region is of the same order as the lifetime of the region, so, in general, pressures will equilibrate. In the Mappings code, for constant-pressure models, the density is a dependent parameter, set by $P$ and $T_e$. The density structure emerges from the models, and the densities vary with the model.  For low metallicity (1/20 solar) and $\log(P/k)$=5, model hydrogen (ionized and neutral) number densities vary typically from 1.5 to 20 cm$^{-3}$ at inner and outer zones.  For higher metallicities (1 solar), hydrogen number  densities vary from 5 to 290 cm$^{-3}$ at inner and outer zones.  For $\log(P/k)$=6, the comparable figures are 15 to 150 cm$^{-3}$ and 40 to 1700 cm$^{-3}$.

Requiring the pressure to be fixed affects the ionization parameter, $\log(q)$, and at higher densities, $\log(q)$ can increase significantly in the inner parts of the \HII region, to $\log(q)$ $>$ 9.3 for an initial model setting (at the inner boundary of the \HII region) of $\log(q)$=8.5, and consequent significantly higher electron temperatures. Figure \ref{tezpk} shows the model results for pressures of $\log(P/k)$=5 and 6, corresponding approximately to electron densities of 5 cm$^{-3}$ and 50 cm$^{-3}$. These models use the same excitation sources as the other models in this work, and therefore the same number and energy distribution of ionizing photons, except where stated, as in  Section 5.  The higher electron density SIGRID objects (5 cm$^{-3} < \rho_e < 100$ cm$^{-3}$), determined from the ratio of the 6716 and 6731\AA~ S~{\sc ii} lines \citepalias[see][]{2014ApJ...786..155N} are shown in blue. They lie above the $\log(P/k)$=5 model curve. It is clear that higher electron densities can explain some apparent electron temperatures above the standard ($\log(P/k)$=5) curve, though not to the extent of the optically thin models. In particular, the measured electron density for the low abundance SIGRD object, J1225-06s2 is $<$ 10 cm$^{-3}$, so its position on the graph is most likely explained by it being optically thin. 

The results for the object HS0837+4717 and several of the emission line galaxy data lie significantly above the model curves, which suggests that other factors such as optical thinness also contribute to generating higher apparent electron temperatures.  This is shown in Figure \ref{tezpkt}, where the magenta curve takes into account both higher densities, corresponding to $\log(P/k)$=6, and optical thickness $\tau$=1.  This is an extreme case, and indicates a likely upper limit for the apparent electron temperatures in the spherically symmetric model  \HII regions, assuming equilibrium electron energies. We will show later that higher electron densities affect the strong line diagnostic grids, and improve the fit to observed data. 

\begin{figure}[htbp]
\centering
\includegraphics[width=0.8\hsize]{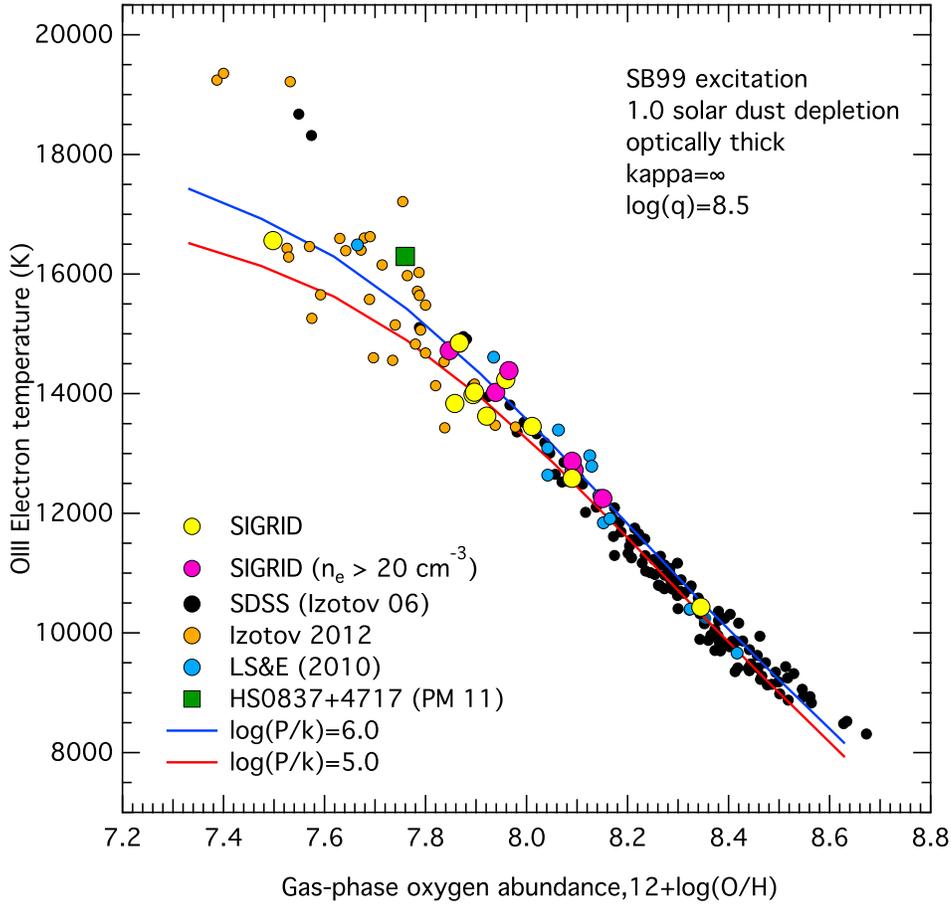}
\caption{Mappings models using different electron densities. The SIGRID objects with measured electron densities n$_e >$ 20 cm$^{-3}$ are marked in magenta.}\label{tezpk}
\end{figure}

\begin{figure}[htbp]
\centering
\includegraphics[width=0.8\hsize]{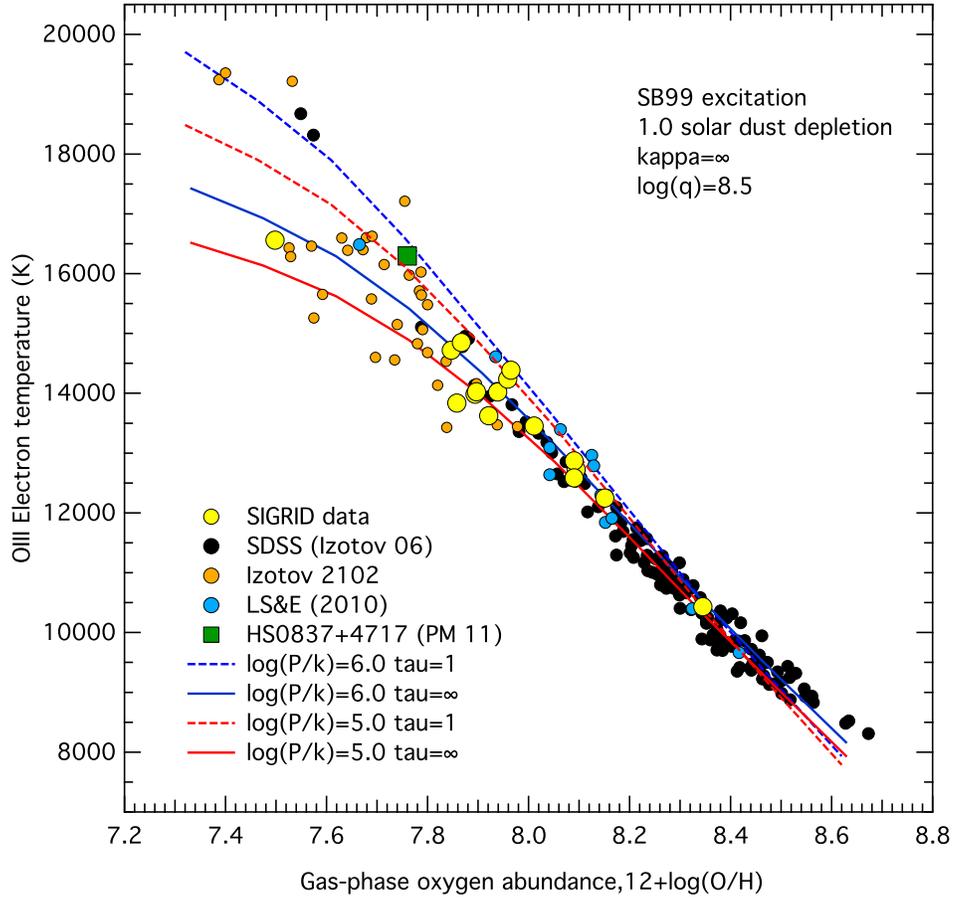}
\caption{As for Figure \ref{tezpk}, but showing optical thicknesses $\tau=\infty$ and 1. The optically thin higher pressure model can account for the observational data.}\label{tezpkt}
\end{figure}

\FloatBarrier

\section{Effect of the ionization parameter, $q$} 

The ionization parameter $q$ is the ratio of the number of ionizing source photons passing through a unit volume to the neutral hydrogen density, as discussed in Section 2.2, above.  The maximum value of $q$ depends on the nature of the central star cluster and the size of the evacuated region, and thus on the strength of the stellar winds from the central cluster.  

The behavior of the ionization parameter, $q$, was analyzed in detail by \cite{2006ApJ...647..244D}. They found that $q$ decreases as an ionizing central star cluster ages: after 2 Myr, $q$ decreases rapidly due to the decrease in ionizing flux as massive stars evolve into red supergiants, and the increase in the mechanical energy due to the stellar winds of Wolf-Rayet stars. At $\sim$3.5 Myr, supernova  explosions contribute further to the internal pressure of the bubble. The ionization parameter depends strongly on the chemical abundance, through two causes. First, at higher abundance, the stellar winds have a higher opacity and absorb a greater fraction of the ionizing photons, thus reducing $q$. Second, stellar atmospheric scattering of photospheric photons is more efficient at higher abundance, leading to more efficient conversion of luminous energy to mechanical energy at the source of the stellar winds, and leading to a reduction in $q$. \cite{2006ApJ...647..244D} found that these factors lead to a dependence of $q$ on oxygen abundance of $q \propto Z^{-0.8}$. This dependence of $q$ on oxygen abundance is illustrated in Figure \ref{tezq}: the model curves at lower $\log(q)$ track the behavior of the SDSS objects better as oxygen abundance increases. This trend was earlier noted for SDSS galaxies by \cite{2006ApJ...647..244D} and for \HII regions in general by \cite{2000ApJ...542..224D}.

A more realistic description (and the subject of future work) would be to plot a single curve where the $\log(q)$ value varies continuously with oxygen abundance. This would more accurately fit the observed data for 12+log(O/H) $\gtrsim$ 8.3.

\begin{figure}[htbp]
\centering
\includegraphics[width=0.8\hsize]{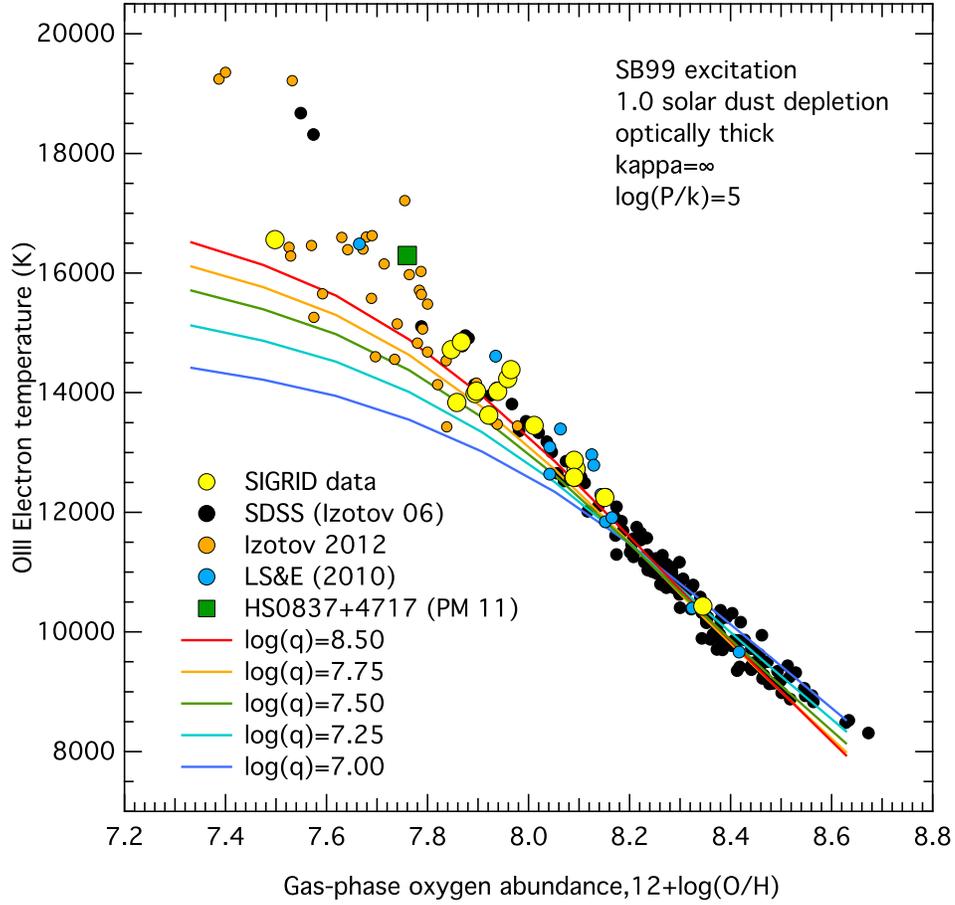}
\caption{Mapplings IV model curves for the [O~{\sc iii}] electron temperature as a function of oxygen abundance, for a range of values of the ionization parameter $\log(q)$. Lower values of $\log(q)$ track the observations better at higher oxygen abundances.}\label{tezq}
\end{figure}

\FloatBarrier

\section{The effect of dust}

As stars in galaxies evolve, older (AGB) stars, WR stars, and supernova ejecta enrich the interstellar gas, and dust forms, locking up oxygen in silicates and carbonaceous grains. This dust is an important component of the molecular clouds where star clusters form that ionize new \HII regions.}  Dust is important in the thermal balance of present day \HII regions \citep[see, e.g.,][]{2003adu..book.....D}. \citet{2011ApJ...729L...3D} find that dust cooling affects the fragmentation of very low-metallicity gas clouds and plays an important role in shaping the stellar IMF even at very low metallicities.

The amount of oxygen present in dust is not readily measurable from the optical spectra, but we estimate it here at $\sim$15\% of the total. This corresponds to a contribution from dust to the total oxygen abundance of 0.07 dex. In this work, we calculate the fraction of other elements in dust grains with respect to oxygen using the abundance ratios from the solar data from \cite{2010Ap&SS.328..179G}. This is discussed in detail in \cite{2013ApJS..208...10D}.

The nature of the dust in \HII regions is poorly understood. \citet{2004MNRAS.350.1330V} studied the effects of dust grain size on the observed features of \HII regions, using models based on the CLOUDY spectral synthesis code. They found that grain size can have a significant effect on the emitted spectrum. Unlike the CLOUDY code, Mappings fully resolves the grain charge for all the grain sizes.\footnote{The energy of photoelectrons emitted by dust grains depends on the charge of the grain. In addition to the work function absorbing the photoionizing energy, the grain charge creates a potential barrier that affects the energy available to the ejected electron, which is then available to heat the surrounding medium. The stronger the initial positive charge of the grain, the less the available heating from the ejected electron. The grain potential is a function of grain size, as their absorption spectrum changes with grain size \citep[e.g.][]{1984ApJ...285...89D} so to compute the photoelectric heating due to dust, a spectrum of grain sizes needs to be calculated and integrated to give the heating. A  single average grain size does not replicate this.} Further, \citet{2004ApJS..153....9G} noted that several of the observable properties of dust can be equally well reproduced by different theoretical grain size distributions and compositions. Thus the observable properties of dust are not especially sensitive to dust grain sizes, but the question obviously requires more work. As a consequence, Mappings uses  the simplest grain composition and grain size distribution model that is consistent with the known properties of the dust in the solar neighborhood, adopting the dust grain size power law model from \citet{1977ApJ...217..425M}. In Mappings, the atomic depletion in dust is consistent with amount of dust formed.

Further, the amount of dust in a galaxy is quite variable. \cite{2014A&A...563A..31R} found that there is a large scatter in the gas-to-dust ratio in star-forming galaxies over a 2-dex range of metallicities.  From FIR Herschel measurements, \cite{2014Natur.505..186F} suggest that dust may be rarer in low metallicity galaxies than previously thought, due to the lack of heavier elements in low metallicity environments and dust destruction occurring in active star forming regions. The latter is suggested by the presence of [Fe~{\sc iii}] lines in the spectra of several of the observational data discussed here. Whatever is the case, the oxygen abundance measured from optical spectra is the gas-phase oxygen abundance, and provides a lower limit for total oxygen. As our models use the total oxygen abundance, the model curves plotted in this paper have had the computed dust-phase oxygen subtracted to permit comparison with the spectral data (a reduction of 0.07 dex in abundance).  Dust is also not uniformly distributed in real nebulae, but in the models we present here, we assume it is sufficiently uniform to allow us to account for the dust physics with a uniform model.

\begin{figure}[htbp]
\centering
\includegraphics[width=0.8\hsize]{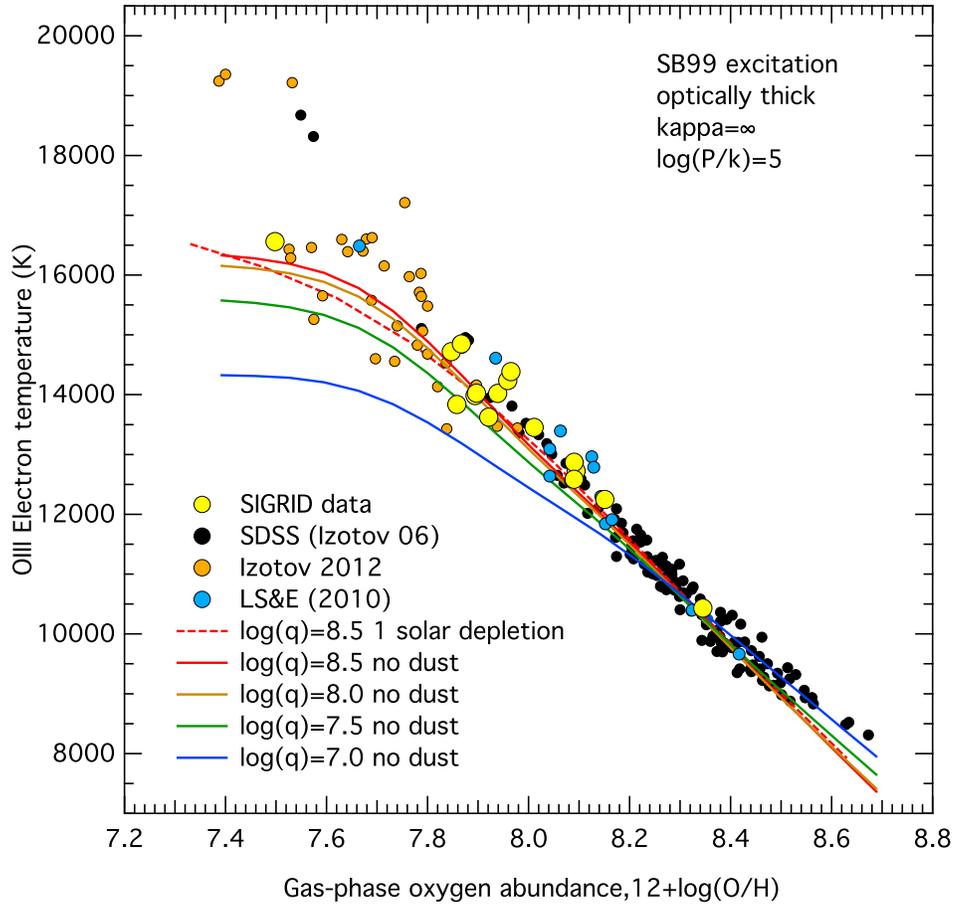}
\caption{Observed [O~{\sc iii}] electron temperature versus gas-phase oxygen abundances, and Mappings photoionization model curves with 1 solar depletion dust, and no dust, for a range of log(q).}\label{tezdu}
\end{figure}

Figure \ref{tezdu} shows the plot from Figure \ref{teza}, with additional curves for dust-free models. Dust depletion appears to cause a small reduction in [O~{\sc iii}] electron temperature between 1/20 and 1/5 solar metallicity, but the net effect is not large.

\FloatBarrier

\section{Effect of non-equilibrium $\kappa$ electron energy distributions}

The effects of non-equilibrium electron energy distributions in \HII regions were explored by \cite{2012ApJ...752..148N,2013ApJS..207...21N}. The current version of the Mappings photoionization code includes the capability to calculate the effect of $\kappa$ electron energies \citep{2013ApJS..208...10D}. $\kappa$-distributions have a high energy power law tail and can readily affect the apparent electron temperature. They can be generated by Alfv\'en waves, magnetic re-connection, shocks, super-thermal atom or ion heating (as in a stellar wind \HII region interaction zone), or by fast primary electrons produced by photoionization with X-ray or EUV photons. These are all sources of long-range injection of high energy electrons, distinct from the normal UV photoionization effects from central star clusters.  

\begin{figure}[htbp]
\centering
\includegraphics[width=0.8\hsize]{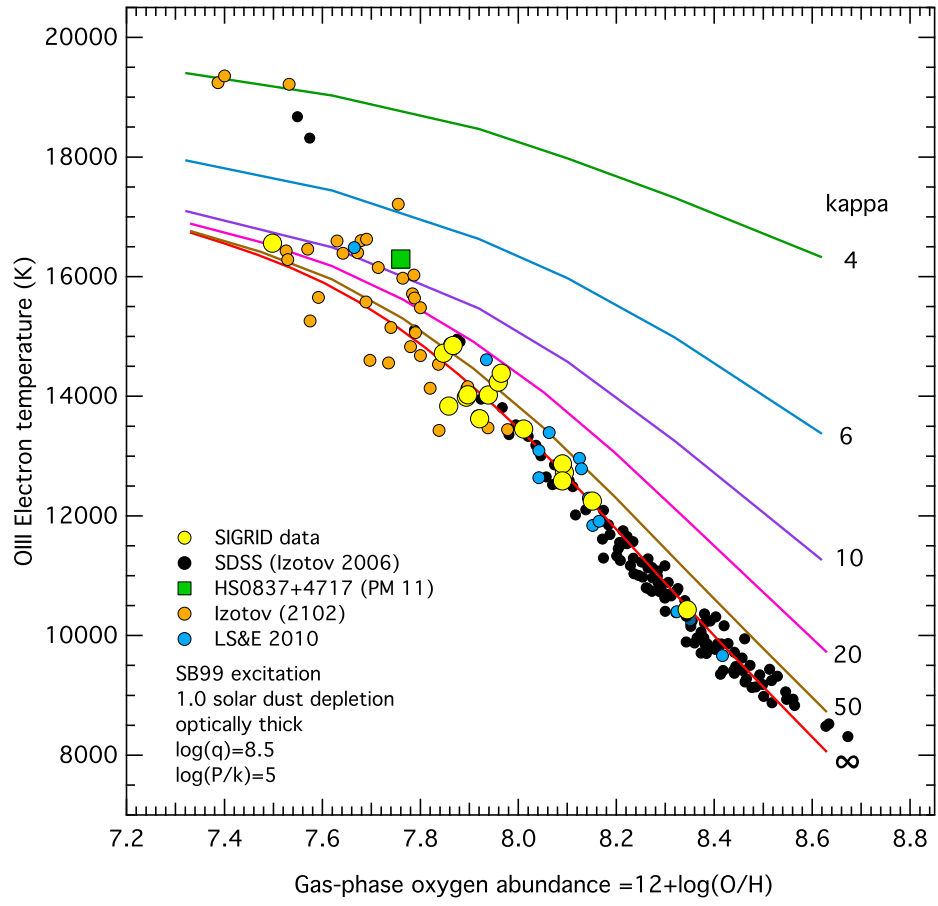}
\caption{Mappings models using different $\kappa$ electron energy distributions.}\label{tezkap}
\end{figure}

Figure \ref{tezkap} shows the result of Mappings models for $\kappa$ values of 4, 6, 10, 20, 50 and infinity (equilibrium), with model curves plotted for total oxygen abundances between 1/20 and 1 solar\footnote{The temperatures and oxygen abundances plotted are the values that one obtains by assuming emission line fluxes from a Maxwell-Boltzmann equilibrium energy distribution \citep{2012ApJ...752..148N}}.  At values of $\kappa >$ 10 (i.e., close to Maxwell-Boltzmann equilibrium), the effects are more obvious for higher metallicities, but for lower, more extreme departures from equilibrium, these electron energy distributions appear capable of describing the excess flux of [O~{\sc iii}] 4363\AA~observed in objects with low oxygen abundance.  As before, we limit the plotted observations to those with apparent electron temperatures  $<$ 20,000K, and oxygen abundance $>$ 1/20 solar, consistent with the limits to the model ranges. It is evident that appropriate values of $\kappa$ can describe the apparent electron temperature behavior, and at face value, $\kappa$ electron energies are an alternative explanation to optically thin nebulae at higher pressures.

However, two important questions arise in interpreting Figure \ref{tezkap}.  The first is, `Can such extreme departures from thermal equilibrium occur in \HII regions?' At present the only direct measurements we have as guides are from solar system plasmas. $\kappa$ energy distributions appear almost universally in solar system plasmas, from the solar wind, through planetary magnetospheres, to the heliosheath.  In particular, proton energy distributions where 1.6 $\lesssim \kappa <$ 2.5 are observed directly in the heliosheath \citep{2011ApJ...741...88L, 2012ApJ...749...11L, 2011ApJ...734....1L}. Using entropy considerations based on Tsallis q--nonextensive statstics \citep{2009insm.book.....T}, these authors argue that stable states can exist that are significantly out of equilibrium. If the entropy-based arguments hold, it is possible that stable extreme $\kappa$ electron energy distributions occur in \HII regions.

The second question, if $\kappa$ electron energy distributions are present,  is, `Why does the effect of more extreme distributions start to appear at oxygen abundances below 0.15 solar (12+log(O/H)$\sim$7.8)?'  Figure \ref{tezkap} suggests that the departure from the equilibrium model (bottom, red) curve commences (for some but not all objects) at this oxygen abundance.  If this departure is caused, or contributed to in significant part, by extreme $\kappa$ distributions, there are evidently physical phenomena affecting the generation of $\kappa$ distributions that come into play at lower metallicities that are less important at higher metallicities.  One may speculate on two possible causes. The first is some form of ``quenching'' of kappa distributions at higher metallicities.  This may not be correct, if the conclusions of \cite{2013ApJS..208...10D} are correct, where they find that values of $\kappa \sim$ 20 explain the behavior of electron temperature for observed data at higher metallicities, $>$ 1 solar. This would imply the presence of $\kappa$ electrons at higher metallicities. Values of $\kappa \sim$ 50 may also explain the trend in electron temperatures for 12+log(O/H) $\gtrsim$ 8.4.

A second possibility is if there are excitation mechanisms present in the stellar winds around low metallicity \HII region star clusters that are less prevalent in stellar winds at higher metallicities.  With our current limited knowledge of the physics of low metallicity O-- and B--star winds, we can say no more than that this explanation is plausible.

Finally, if $\kappa$ distributions are present in parts of \HII regions, the effect on the spectrum of averaging over the whole \HII region, such as we have with the observed data here, may dilute the observed non-equilibrium effects.  The spatial variation in $\kappa$ distributions, if present, is likely to be more obvious in nearby \HII regions, where individual structures in the nebulae can be distinguished. 

\FloatBarrier

\section{Summary of factors affecting the measurement of apparent electron temperature}

Using the Mappings photoionization code and a three-dimensional Str\"omgren Sphere model, we find that the `standard' equilibrium model ($\log(q)=8.5$, $\tau=\infty$, $\kappa=\infty$, Figure \ref{teza} and Equation \ref{eq7}) does a good job of fitting the majority of the observational data points for oxygen abundances $>$ 1/5 solar. There are a number of parameters that can be used to explain divergences in temperature from the equilibrium model for lower metallicities. Higher temperature objects that lie above the `standard' curve can be explained by electron density variations, higher energy excitation sources, dust, optically thin nebulae, and non-equilibrium $\kappa$ electron energy distributions.  Low optical depth appears to make an important contribution, and its contribution can be confirmed using diagnostics such as the [S{\sc ii}]/[O{\sc iii}] line ratio (Figure \ref{s2o3}).  Electron densities and high energy excitation sources also contribute to higher electron temperatures. Dust appears not to make a major contribution, if it is uniformly distributed. There is a considerable degree of degeneracy between these factors, but the equilibrium model `standard' curve provides a reliable first order basis for determining total oxygen gas-phase abundance for a given measured O~{\sc iii} apparent electron temperature, above 1/15 solar abundance. The error bars on the data are in some places larger than the trends the data exhibit, but the trends nonetheless illustrate the effects of the physical parameters we have discussed. For oxygen abundances below 1/15 solar in the spherical \HII region model, high apparent electron temperatures can be explained by non-equilibrium electron energy distributions, or a combination of higher pressure and low optical depth. The former requires the additional explanation of why its effects are only apparent for oxygen abundances below 0.15 solar.

Finally, the photoionization model assumptions are important. Mappings is a comprehensive modeling system, which uses the current best available atomic data.  The code has been revised in a few key areas since the earlier paper \citep{2013ApJS..208...10D} to give more consistent results (see appendix), but the numerical results are not substantially changed, with the exception of somewhat higher [O~{\sc iii}] apparent electron temperatures at low oxygen abundance (+$\sim$500K)\footnote{We plan to make Mappings IV available as a web page application for real-time use in the near future.}.  The formulae used here and in \citetalias{2014ApJ...786..155N} are those derived from the earlier paper.  Errors arsing from collision strength uncertainties may amount to a difference in gas-phase oxygen abundance of $\sim\pm$0.03, leading to uncertainties in the model curves that are within the published uncertainties in the data. The most significant unknown is how well a symmetric Str\"omgren Sphere model is able to replicate the behavior of real \HII regions.  We plan to explore this further.

\section{Diagnostic grids}

\subsection{Two-dimensional grids for different  densities and Lyman limit optical depths}

The convention in strong line methods is to plot observed data values on a two-dimensional grid of line ratios that will allow measurement of metallicities \citep[e.g.,][]{2002ApJS..142...35K, 2008ApJ...681.1183K}. \cite{2013ApJS..208...10D} presented new diagnostic strong line grids computed using the Mappings code, for a range of values of oxygen abundance, 12+log(O/H), the ionization parameter, $\log(q)$, and the non-equilibrium energy distribution parameter, $\kappa$, for optically thick nebulae at low density.  As shown in \citetalias{2014ApJ...786..155N}, some of the SIGRID data points did not fit on these standard diagnostic grids, meaning it was not possible to use the grids to measure abundances and ionization parameters.  This led us to ask if some variation in the grids would permit the measurement of 12+log(O/H) and $\log(q)$ for objects that did not fit on the standard grids. Fortunately, it is a simple matter to compute grids with different physical parameters.  Here we present grids for higher pressure ($\log(P/k)$=6) than the earlier published grids, and a range of values of the optical depth, $\tau$ and the non-equilibrium energy parameter, $\kappa$.

\begin{figure}[htbp]
\centering
\includegraphics[width=1.0\hsize]{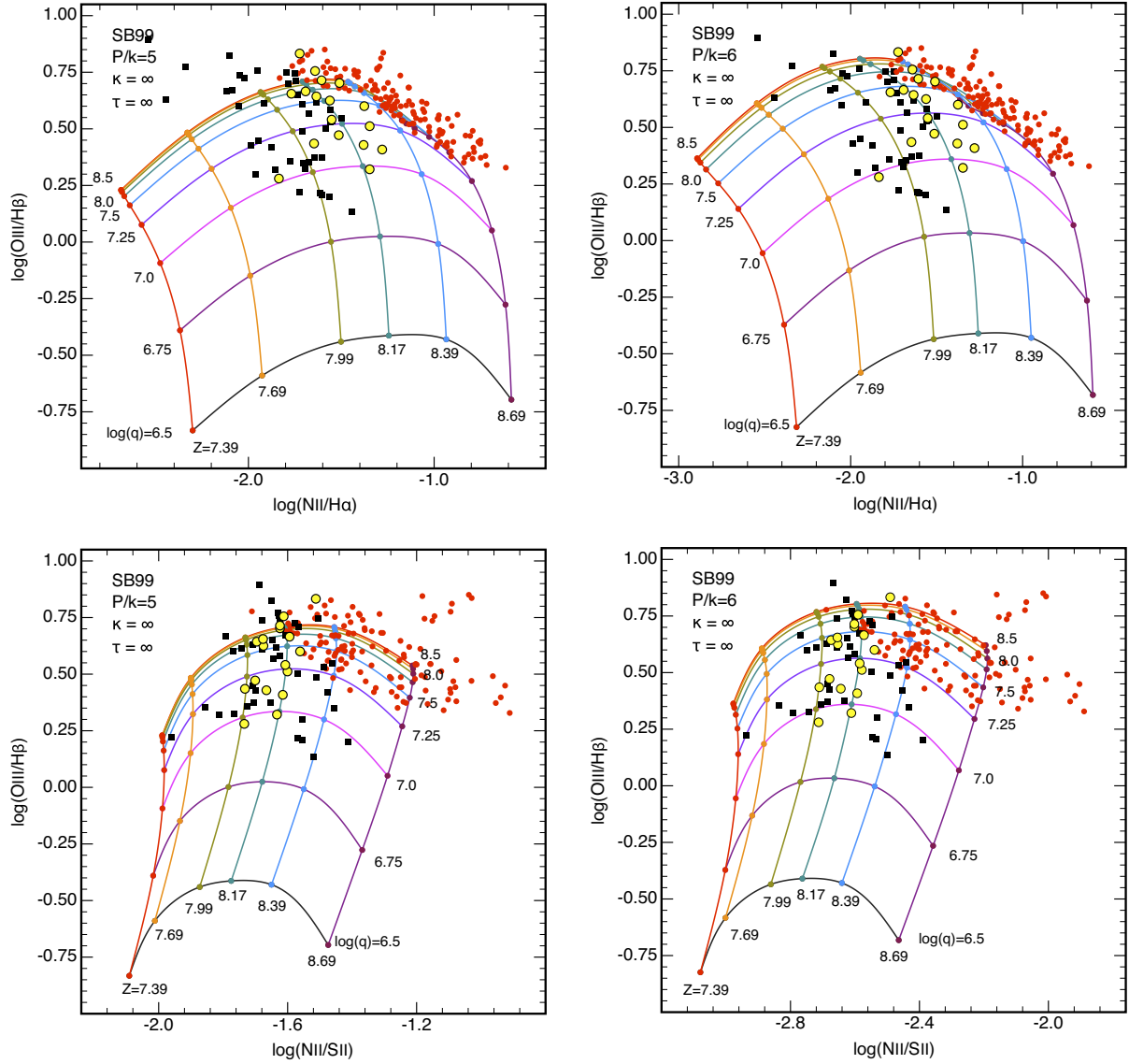}
\caption{Observational data (SIGRID, yellow; Izotov et al. (2006), red; and Izotov et al. (2012), black) plotted on diagnostic grids for log(O{\sc iii}/H$\beta$) versus log(N{\sc ii}/H$\alpha$), upper row, and log(O{\sc iii}/H$\beta$) versus log(N{\sc ii}/S{\sc ii}), lower row, for low density, $\log(P/k)$=5 (left column) and higher density, $\log(P/k)$=6 (right column). The error bars for the SIGRID objects are given in Figures 10 and 12 of \citetalias{2014ApJ...786..155N}. The errors involving [N{\sc ii}] fluxes are significantly greater than those for [O{\sc iii}]. They have been omitted here to aid clarity.}\label{pk56}
\end{figure}

Figure \ref{pk56} compares two diagnostic grids, log([O{\sc iii}]/H$\beta$) versus log([N{\sc ii}]/H$\alpha$) (upper row), and log([O{\sc iii}]/H$\beta$) versus log(N{\sc ii}/S{\sc ii}) (lower row) for the standard lower pressure (and thus density) case ($\log(P/k)$=5, left column), and at higher pressure ($\log(P/k)$=6, right column). All but one SIGRID object can be accommodated on the grids for $\log(P/k)$=6. This is consistent with Figure \ref{tezpk}, where the higher density objects tend to lie above the standard T$_e$ versus 12+log(O/H) curve. The grids give similar values for oxygen abundance for either value of $\log(P/k)$ for each observation, but the higher $\log(P/k)$ set give slightly lower values for $\log(q)$.  This can be understood from Figures \ref{tezpk} and \ref{tezq}---the boost in $\log(P/k)$ is compensated for by a drop in $\log(q)$, which brings down the $\log(q)$ values for outliers onto the diagnostic grid.

Figure \ref{pk56} also shows two additional data sets.  The red points, from \cite{2006A&A...448..955I}, show scatter, probably caused by measurement uncertainties.  The abundance values between the two sets of grids (upper and lower rows) are also somewhat discrepant.  This is most likely due to uncertainties in the measurements, the de-reddening, and in the collision strength data used to create the grids.  However, the abundances are in general much more consistent than the older strong line methods.  Of the two diagnostics, the lower row is least susceptible to de-reddening errors and is the more reliable, as well as avoiding the degeneracy apparent in the upper row for higher ionization parameters and abundances.

\FloatBarrier

\begin{figure}[h]
\includegraphics[width=0.5\hsize]{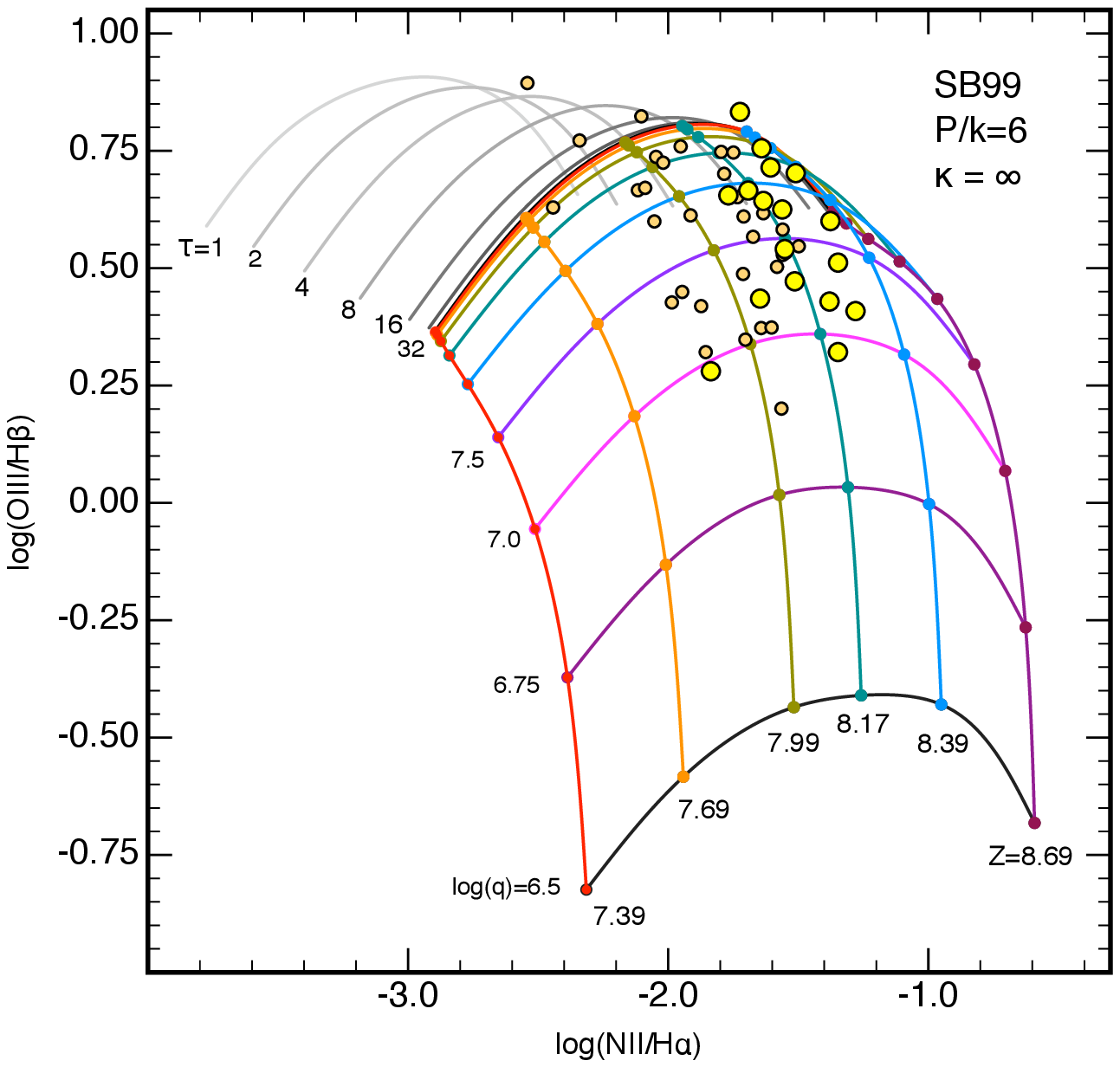}
\includegraphics[width=0.5\hsize]{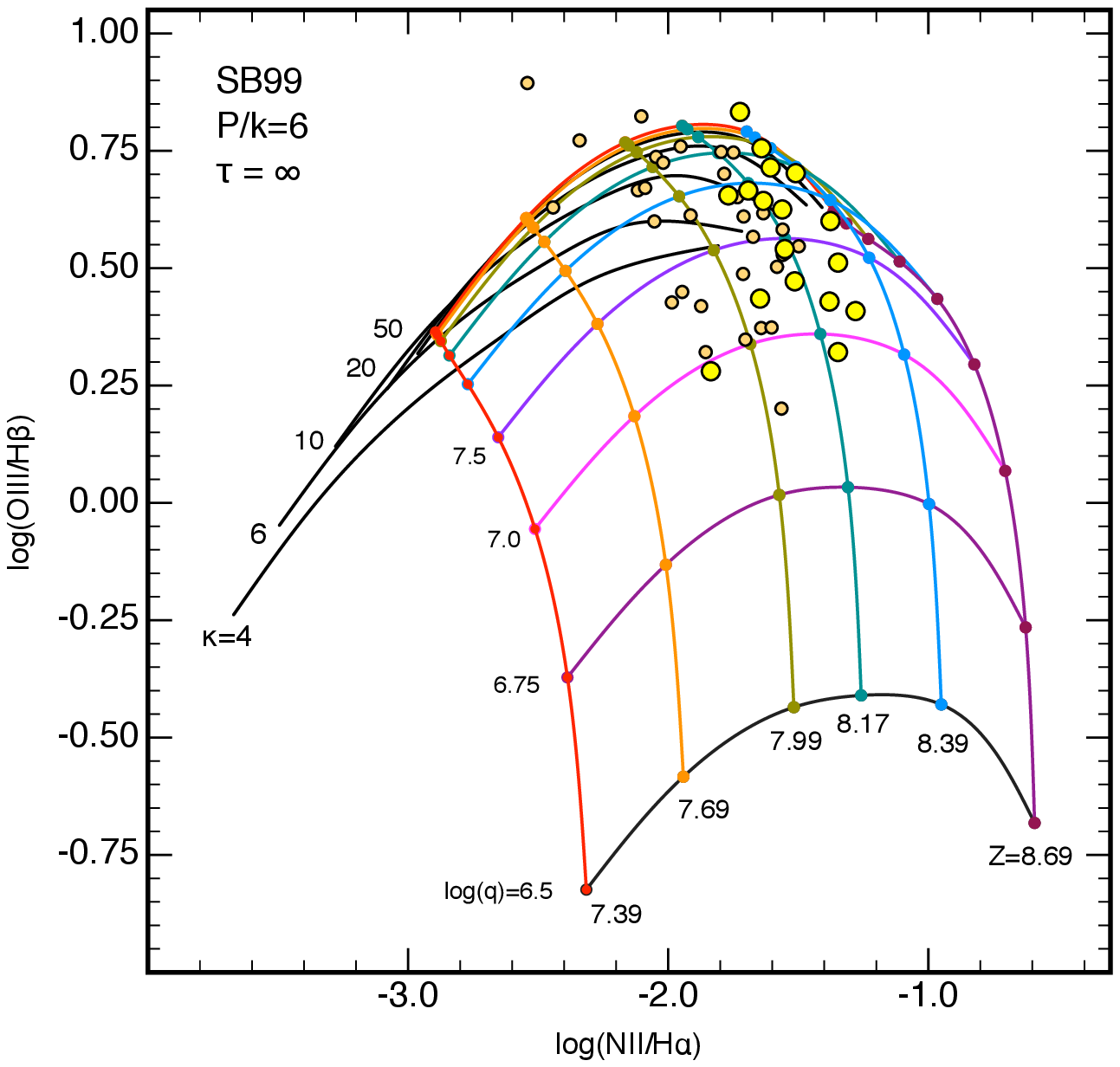}
\caption{SIGRID data plotted on diagnostic grids for log(O{\sc iii}/H$\beta$) versus log(N{\sc ii}/H$\alpha$), overlaid with plots for a range of values of optical thickness, $\tau$=1, 2, 4, and 8, 16, 32 at $\log(q)$=8.5 (left pane), and for $\kappa$= 4, 6, 10, 20, and 50 at $\log(q)$=8.5. The latter curves fold behind the basic plot.}\label{pk5kt}
\end{figure}

Figure \ref{pk5kt} shows the grids for $\log(P/k)$=6 (as in Figure \ref{pk56}, upper right panel) with additional curves for a range of optical depths (left panel), and a range of values of $\kappa$ (right panel).  It suggests that lower optical depths go some way to explaining the off-grid data points, whereas grids involving lower $\kappa$ values do not significantly improve the fit of the observations onto the grids. In both cases, the $\tau$-- and $\kappa$--extended grids roll away from the data points, in the case of $\kappa$, very rapidly, so that this diagnostic plot is degenerate for $\kappa$.  So $\kappa$ does not provide information that will allow us to evaluate $12+\log(O/H)$ or $\log(q)$ for points initially off the grid. Both panels suggest three-dimensional shapes for the grids, and that plotting three diagnostic ratios on three orthogonal axes may provide new information. We explore this below.

\FloatBarrier

\section{Investigating three-dimensional diagnostic graphs}

The three-dimensional nature of the diagnostics is suggested in the two dimensional grids, particularly in Figure \ref{pk5kt}.  This invites us to plot three-dimensional figures to explore the possibility further. Plotting and manipulation of three-dimensional graphs is not simple.  Vogt et al. (in press) have used Python routines to generate two-dimensional projections of three-dimensional graphs for high metallicity objects from the SDSS data, allowing them to separate \HII regions from AGNs and LINERs.  For this work we have used the GRAF application written for OS X by one of the present authors (RSS), which provides a particularly flexible tool for investigating the three-dimensionality of the diagnostics, although we have not yet used it to generate 3-D PDF files\footnote{~3-dimensional projected diagrams displayed here were prepared using GRAF 4.7.3 (OS X) software authored by R.S.S., available as (unsupported) beta software at http://miocene.anu.edu.au/graf/}.

\begin{figure}[h]
\includegraphics[width=0.5\hsize]{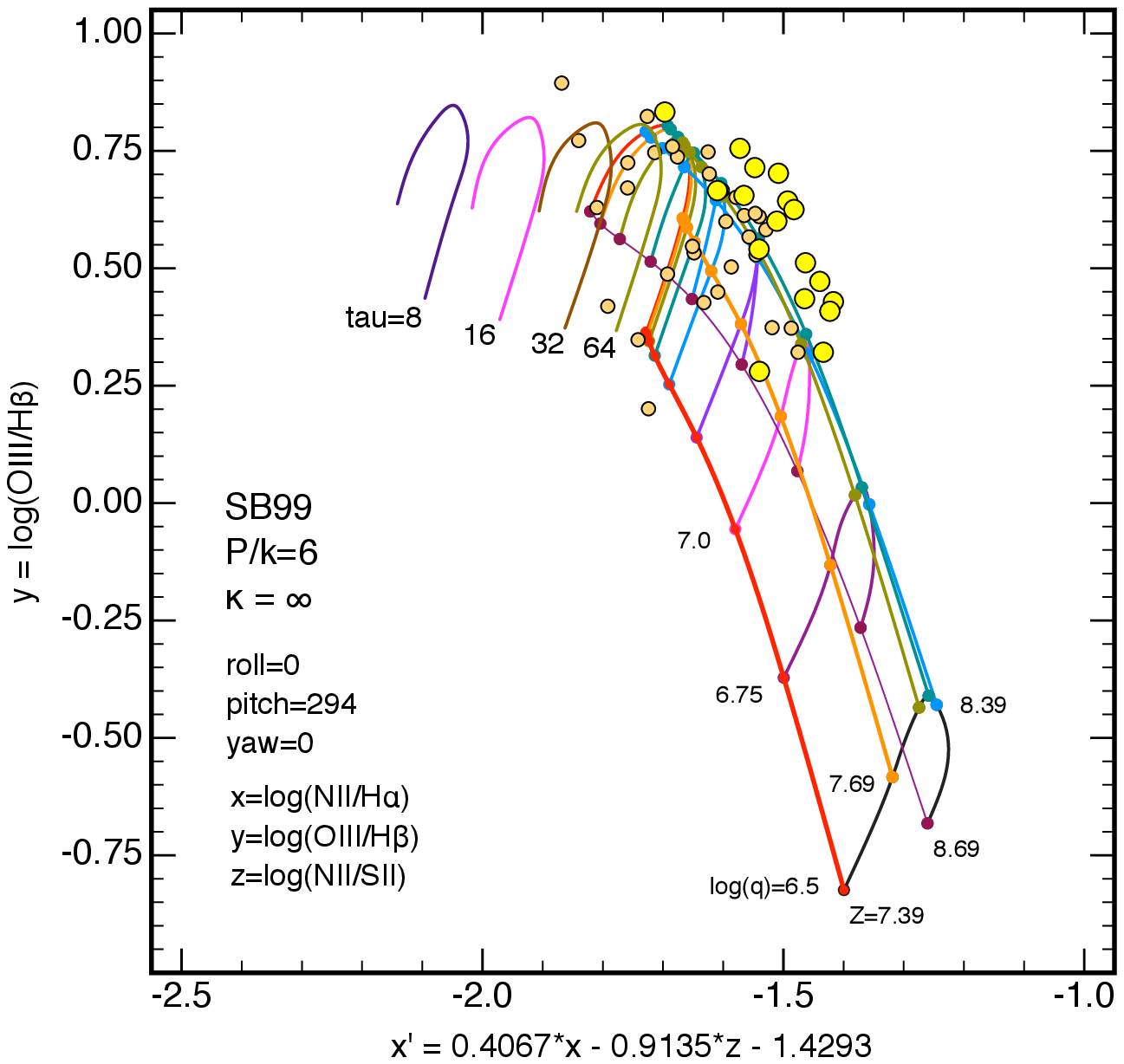}
\includegraphics[width=0.5\hsize]{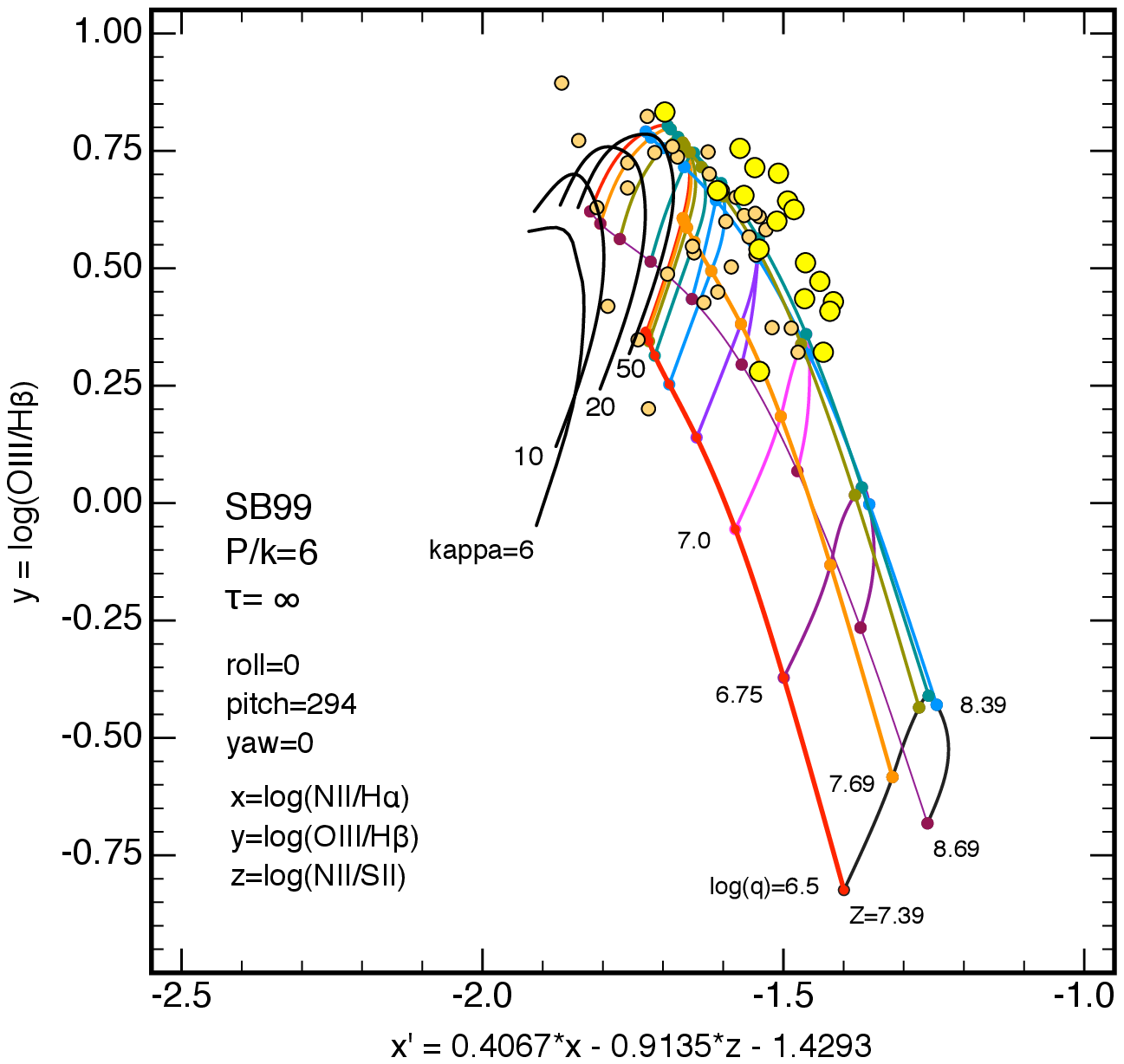}
\caption{Left Panel: Three-dimensional projection of log(N~{\sc ii}/H$\alpha$) and log(N~{\sc ii}/S~{\sc ii}) versus log(O~{\sc iii}/H$\beta$) diagnostic grids (Figures \ref{pk56}, \ref{pk5kt}, with $\log(P/k)$=6), rotated about the y-axis as indicated, showing the optically thick equilibrium diagnostic grids with the $\log(q)$=8.50 grid lines for four values of the \HI optical depth, $\tau$. Right Panel: Similar grids, but with the $\log(q)$=8.5 grid lines for four values of the non-equilibrium parameter $\kappa$. Both figures show the data for the SIGRID objects (larger circles) and from \citet{2012A&A...546A.122I} (smaller circles).}\label{p412}
\end{figure}

Figure \ref{p412} shows the outcome of this three-dimensional manipulation.  We have combined three diagnostic ratios for log(N{\sc ii}/H$\alpha$), log(O{\sc iii}/H$\beta$) and log(N{\sc ii}/S{\sc ii}) on the $x$, $y$ and $z$ axes respectively, and have rotated these about the y-axis to compare the location of the observational data with respect to the diagnostic curves.

The main grid in both panels is for a range of oxygen abundance between 1/20 solar and 1 solar, and a range of ionization parameters between 6.5 and 8.5 (log), for $\log(P/k)$=6, $\kappa=\infty$ and $\tau=\infty$. The additional free standing arcs are the grid lines for the same abundance range and $P/k$, for $\log(q)$=8.5, for $\kappa=\infty$ and $\tau$=8, 16, 32 and 64 (left panel); and for $\tau=\infty$ and $\kappa$=6, 10 20 and 50 (right panel). We have only plotted the highest value of $\log(q)$ for the non-equilibrium and optically thin cases, to make the plots easier to read.
 
The data from SIGRID observations and from \citet{2012A&A...546A.122I} are plotted over these curves.  It is evident that the observational data  are described better by the left panel (optically thin and equilibrium energies) than by the right (optically thick and $\kappa$ electron energies).  This illustrates how three-dimensional diagnostics may be used to discriminate between two otherwise alternative two-dimensional descriptions.

While these results are preliminary in nature, they demonstrate the utility of three-dimensional manipulation of three sets of diagnostic ratios in augmenting the standard two-dimensional grids, to show which parameters can be used to fit the observations better to the grids. Although we have not been able to demonstrate this in the present paper, manually rotating the diagrams shows that the location of several of the observed data points are better described by optically thin diagnostic grids than by the non-equilibrium grids.

\FloatBarrier 

\section{Discussion}

Measured [O~{\sc iii}] electron temperatures reported in the literature using simple analysis techniques only provide general guides to the physical conditions in an \HII region.  No \HII region is thermally uniform, so temperatures measured from the emitted spectra can only be a volume and intensity weighted  average of the internal conditions.  In addition to this averaging process, other parameters impact on the measured spectra.  These include the electron and hydrogen densities, the nature of the central ionizing radiation source, whether the electron energies are in equilibrium or not, and possibly most important, whether the nebula is optically thin.

Increasing attention has been paid in recent years to the possibility of optically thin nebulae \citep[][and references therein]{2012ApJ...755...40P}. \citet{2013ApJ...779...76Z}, for example, find that \HII regions in their galaxy sample show evidence of optically thin pathways, provided the systems are old enough ($\gtrsim$3 Myr) for mechanical feedback to have cleared optical thin paths, but young enough ($\lesssim$5 Myr) that the ionizing central star cluster is still present. It appears likely that \HII regions may be optically thin much earlier than this.  In the case of the central UC1 cluster in M17, \cite{1997ApJ...489..698H} studied 9 O-stars still on the zero-age main sequence, suggesting an age for the cluster and the \HII region of 1 Myr or less.  Using JHK band photometry,  \cite{2002ApJ...577..245J} deduced an age for the central cluster of $<$3 Myr. In the same cluster, discrete X-ray sources are present corresponding to individual stars, as well as an extended diffuse X-ray field \citep{2003ApJ...593..874T}.  This ionizing radiation is evidently escaping from the \HII region, indicating the region is optically thin.  This was also reported for the ultra-low abundance object, I Zw 18, by \citet{2008A&A...478..371P}. Given the \HII regions are inhomogeneous, it is plausible that parts of them may be optically thin as soon as they are visible.  So it is to be expected that some of the observations considered here exhibit characteristics readily explained by low values of the optical depth, $\tau$.  As we note, the radiation emitted  from our optically thick model nebulae is brighter than from otherwise similar optically thin nebulae.  For a distant \HII region, given the likelihood that some of a nebula will be optically thick and other parts optically thin, these regions will not be resolved, and the resultant spectrum is a weighted mean of the two.  The summed or `hybrid' result is skewed in favor of the brighter, optically thick contribution. Thus any suggestion of optical thinness in the integrated spectra indicates a substantial fraction of the nebula is optically thin. The effect on apparent electron temperatures of these hybrid spectra are shown in Figure \ref{taurat}.

The next most important parameter affecting how well the model T$_e$ versus $12+\log(O/H)$ curves fit the observations is $\log(P/k)$.  The SIGRID observations for objects whose electron densities, determined from the S~{\sc ii} line ratios, are $>$20 cm$^{-3}$ lie consistently above the standard model ($\log(q)$ = 8.5, $\tau=\infty$, $\kappa=\infty$, $\log(P/k)$ = 5), and their positions on the plot are therefore readily explained by higher pressure and consequent higher electron densities (Figure \ref{tezpk}). 

The nature of the UV spectral energy distribution of the ionizing star cluster is also of critical importance to the resulting nebular spectra, but it is one where the modeling process is probably least well developed.  The photoionization models depend on there being reliable stellar spectral energy distribution models, and at low metallicities extensive stellar models  are lacking.  We use the Starburst99 models that give reasonable high energy photon fluxes, but we are limited to a minimum oxygen abundance of 0.05 solar.  This is an area where further work is important, to understand the behavior of low metallicity \HII regions affected by strong starburst events. The data considered here are consistent with the findings of  \cite{2006ApJ...647..244D} that the ionization parameter decreases with increasing metallicity.  Finally, our models suggest that $\kappa$ non-equilibrium electron energy distributions can generate high apparent electron temperatures, but if this is correct, it is not clear why their effect only becomes apparent at low metallicity. The three-dimensional diagnostic grid plots suggest that $\kappa$ is a less satisfactory explanation for these high electron temperatures than a combination of low optical depth and higher pressure.

\section{Conclusions}

In this paper we have explored the various parameters affecting the measurement of apparent electron temperatures  in \HII regions. We have used the data for the SIGRID observations from \citetalias{2014ApJ...786..155N}, the SDSS data from \cite{2006A&A...448..955I}, low metallicity emission line galaxy data from \citet{2012A&A...546A.122I} and for Wolf-Rayet galaxies from \citet{2009A&A...508..615L,2010A&A...517A..85L} to demonstrate these effects.  The standard equilibrium, optically thick model with $\log(q)$=8.5 fits the observations remarkably well.  The deviations from the fit are slight for metallicities above 1/5 solar, and can be explained by other physical parameters coming into play. For the low metallicity objects considered here, the important parameters are the ionization parameter, $\log(q)$, the optical thickness, $\tau$, $\kappa$ electron energies, and the pressure/density, specified by $\log(P/k)$.  The nature of the central ionization star cluster source is also important, but the models available for spectral energy distributions at low metallicity limit our exploration here. This is an important area for further investigation.

We have developed diagnostic plots to identify objects that are not optically thick, and it is clear that some of the observations can be readily explained by this effect.  Other objects show clear evidence of the effects of higher electron densities.  Together, these two factors offer a plausible explanation of the observed data. While non-equilibrium $\kappa$ electron energy distributions may affect the observations, they require very low values of $\kappa$ in some cases, and an explanation for their disappearance at higher abundances.

We have used modeling of three-dimensional grids to augment the standard two-dimensional diagnostics, and suggest that this is an area with considerable potential, given the right manipulative tools.  We conclude that the highest measured electron temperatures that fall within the bounds of the fitting approximations can be satisfactorily explained by a combination of pressure ($\log(P/k) \gtrsim$ 6) and optically thin nebulae (8 $< \tau < \infty$).

\begin{acknowledgments}
Mike Dopita acknowledges the support of the Australian Research Council (ARC) through Discovery  project DP0984657. This work was funded in part by the Deanship of Scientific Research (DSR), King Abdulaziz University, under grant No. (5-130/1433 HiCi). The authors acknowledge this financial support from KAU.
\end{acknowledgments}

\bibliographystyle{aa}

\begin{thebibliography}{69}
\expandafter\ifx\csname natexlab\endcsname\relax\def\natexlab#1{#1}\fi

\bibitem[{{Aller}(1984)}]{1984ASSL..112.....A}
{Aller}, L.~H., ed. 1984, Astrophysics and Space Science Library, Vol. 112,
  {Physics of thermal gaseous nebulae}

\bibitem[{{Bland-Hawthorn} {et~al.}(2011){Bland-Hawthorn}, {Sutherland}, \&
  {Karlsson}}]{2011EAS....48..397B}
{Bland-Hawthorn}, J., {Sutherland}, R., \& {Karlsson}, T. 2011, in EAS
  Publications Series, Vol.~48, EAS Publications Series, ed. M.~{Koleva},
  P.~{Prugniel}, \& I.~{Vauglin}, 397--404

\bibitem[{{Bresolin} {et~al.}(2005){Bresolin}, {Schaerer}, {Gonz{\'a}lez
  Delgado}, \& {Stasi{\'n}ska}}]{2005A&A...441..981B}
{Bresolin}, F., {Schaerer}, D., {Gonz{\'a}lez Delgado}, R.~M., \&
  {Stasi{\'n}ska}, G. 2005, \aap, 441, 981

\bibitem[{{Diaz} {et~al.}(1987){Diaz}, {Terlevich}, {Pagel}, {Vilchez}, \&
  {Edmunds}}]{1987MNRAS.226...19D}
{Diaz}, A.~I., {Terlevich}, E., {Pagel}, B.~E.~J., {Vilchez}, J.~M., \&
  {Edmunds}, M.~G. 1987, \mnras, 226, 19

\bibitem[{{Dopcke} {et~al.}(2011){Dopcke}, {Glover}, {Clark}, \&
  {Klessen}}]{2011ApJ...729L...3D}
{Dopcke}, G., {Glover}, S.~C.~O., {Clark}, P.~C., \& {Klessen}, R.~S. 2011,
  \apjl, 729, L3+

\bibitem[{{Dopita} {et~al.}(2006){Dopita}, {Fischera}, {Sutherland}, {Kewley},
  {Tuffs}, {Popescu}, {van Breugel}, {Groves}, \&
  {Leitherer}}]{2006ApJ...647..244D}
{Dopita}, M.~A., {Fischera}, J., {Sutherland}, R.~S., {et~al.} 2006, \apj, 647,
  244

\bibitem[{{Dopita} {et~al.}(2000){Dopita}, {Kewley}, {Heisler}, \&
  {Sutherland}}]{2000ApJ...542..224D}
{Dopita}, M.~A., {Kewley}, L.~J., {Heisler}, C.~A., \& {Sutherland}, R.~S.
  2000, \apj, 542, 224

\bibitem[{{Dopita} \& {Sutherland}(2003)}]{2003adu..book.....D}
{Dopita}, M.~A. \& {Sutherland}, R.~S. 2003, {Astrophysics of the diffuse
  universe} (Springer Verlag)

\bibitem[{{Dopita} {et~al.}(2013){Dopita}, {Sutherland}, {Nicholls}, {Kewley},
  \& {Vogt}}]{2013ApJS..208...10D}
{Dopita}, M.~A., {Sutherland}, R.~S., {Nicholls}, D.~C., {Kewley}, L.~J., \&
  {Vogt}, F.~P.~A. 2013, \apjs, 208, 10

\bibitem[{{Draine} \& {Lee}(1984)}]{1984ApJ...285...89D}
{Draine}, B.~T. \& {Lee}, H.~M. 1984, \apj, 285, 89

\bibitem[{{Esteban} {et~al.}(2009){Esteban}, {Bresolin}, {Peimbert},
  {Garc{\'{\i}}a-Rojas}, {Peimbert}, \& {Mesa-Delgado}}]{2009ApJ...700..654E}
{Esteban}, C., {Bresolin}, F., {Peimbert}, M., {et~al.} 2009, \apj, 700, 654

\bibitem[{{Esteban} \& {Peimbert}(1995)}]{1995A&A...300...78E}
{Esteban}, C. \& {Peimbert}, M. 1995, \aap, 300, 78

\bibitem[{{Esteban} {et~al.}(2004){Esteban}, {Peimbert}, {Garc{\'{\i}}a-Rojas},
  {Ruiz}, {Peimbert}, \& {Rodr{\'{\i}}guez}}]{2004MNRAS.355..229E}
{Esteban}, C., {Peimbert}, M., {Garc{\'{\i}}a-Rojas}, J., {et~al.} 2004,
  \mnras, 355, 229

\bibitem[{{Fisher} {et~al.}(2014){Fisher}, {Bolatto}, {Herrera-Camus},
  {Draine}, {Donaldson}, {Walter}, {Sandstrom}, {Leroy}, {Cannon}, \&
  {Gordon}}]{2014Natur.505..186F}
{Fisher}, D.~B., {Bolatto}, A.~D., {Herrera-Camus}, R., {et~al.} 2014, \nat,
  505, 186

\bibitem[{{Garnett}(1992)}]{1992AJ....103.1330G}
{Garnett}, D.~R. 1992, \aj, 103, 1330

\bibitem[{{Grevesse} {et~al.}(2010){Grevesse}, {Asplund}, {Sauval}, \&
  {Scott}}]{2010Ap&SS.328..179G}
{Grevesse}, N., {Asplund}, M., {Sauval}, A.~J., \& {Scott}, P. 2010, \apss,
  328, 179

\bibitem[{{Groves} {et~al.}(2004){Groves}, {Dopita}, \&
  {Sutherland}}]{2004ApJS..153....9G}
{Groves}, B.~A., {Dopita}, M.~A., \& {Sutherland}, R.~S. 2004, \apjs, 153, 9

\bibitem[{{Hanson} {et~al.}(1997){Hanson}, {Howarth}, \&
  {Conti}}]{1997ApJ...489..698H}
{Hanson}, M.~M., {Howarth}, I.~D., \& {Conti}, P.~S. 1997, \apj, 489, 698

\bibitem[{{Izotov} {et~al.}(2006){Izotov}, {Stasi{\'n}ska}, {Meynet}, {Guseva},
  \& {Thuan}}]{2006A&A...448..955I}
{Izotov}, Y.~I., {Stasi{\'n}ska}, G., {Meynet}, G., {Guseva}, N.~G., \&
  {Thuan}, T.~X. 2006, \aap, 448, 955

\bibitem[{{Izotov} {et~al.}(2012){Izotov}, {Thuan}, \&
  {Guseva}}]{2012A&A...546A.122I}
{Izotov}, Y.~I., {Thuan}, T.~X., \& {Guseva}, N.~G. 2012, \aap, 546, A122

\bibitem[{{Izotov} {et~al.}(1994){Izotov}, {Thuan}, \&
  {Lipovetsky}}]{1994ApJ...435..647I}
{Izotov}, Y.~I., {Thuan}, T.~X., \& {Lipovetsky}, V.~A. 1994, \apj, 435, 647

\bibitem[{{Jiang} {et~al.}(2002){Jiang}, {Yao}, {Yang}, {Ando}, {Kato},
  {Kawai}, {Kurita}, {Nagata}, {Nagayama}, {Nakajima}, {Nagashima}, {Sato},
  {Tamura}, {Nakaya}, \& {Sugitani}}]{2002ApJ...577..245J}
{Jiang}, Z., {Yao}, Y., {Yang}, J., {et~al.} 2002, \apj, 577, 245

\bibitem[{{Kennicutt} {et~al.}(2003){Kennicutt}, {Bresolin}, \&
  {Garnett}}]{2003ApJ...591..801K}
{Kennicutt}, Jr., R.~C., {Bresolin}, F., \& {Garnett}, D.~R. 2003, \apj, 591,
  801

\bibitem[{{Kewley} \& {Dopita}(2002)}]{2002ApJS..142...35K}
{Kewley}, L.~J. \& {Dopita}, M.~A. 2002, \apjs, 142, 35

\bibitem[{{Kewley} \& {Ellison}(2008)}]{2008ApJ...681.1183K}
{Kewley}, L.~J. \& {Ellison}, S.~L. 2008, \apj, 681, 1183

\bibitem[{{Kisielius} {et~al.}(2009){Kisielius}, {Storey}, {Ferland}, \&
  {Keenan}}]{2009MNRAS.397..903K}
{Kisielius}, R., {Storey}, P.~J., {Ferland}, G.~J., \& {Keenan}, F.~P. 2009,
  \mnras, 397, 903

\bibitem[{{K\"oppen}(1979)}]{1979A&AS...35..111K}
{K\"oppen}, J. 1979, \aaps, 35, 111

\bibitem[{{Leitherer}(2014)}]{2014arXiv1402.0824L}
{Leitherer}, C. 2014, ArXiv e-prints

\bibitem[{{Leitherer} {et~al.}(2014){Leitherer}, {Ekstrom}, {Meynet},
  {Schaerer}, {Agienko}, \& {Levesque}}]{2014arXiv1403.5444L}
{Leitherer}, C., {Ekstrom}, S., {Meynet}, G., {et~al.} 2014, ArXiv e-prints

\bibitem[{{Leitherer} {et~al.}(1999){Leitherer}, {Schaerer}, {Goldader},
  {Gonz{\'a}lez Delgado}, {Robert}, {Kune}, {de Mello}, {Devost}, \&
  {Heckman}}]{1999ApJS..123....3L}
{Leitherer}, C., {Schaerer}, D., {Goldader}, J.~D., {et~al.} 1999, \apjs, 123,
  3

\bibitem[{{Lennon} \& {Burke}(1994)}]{1994A&AS..103..273L}
{Lennon}, D.~J. \& {Burke}, V.~M. 1994, \aaps, 103, 273

\bibitem[{{Livadiotis} \& {McComas}(2011)}]{2011ApJ...741...88L}
{Livadiotis}, G. \& {McComas}, D.~J. 2011, \apj, 741, 88

\bibitem[{{Livadiotis} \& {McComas}(2012)}]{2012ApJ...749...11L}
{Livadiotis}, G. \& {McComas}, D.~J. 2012, \apj, 749, 11

\bibitem[{{Livadiotis} {et~al.}(2011){Livadiotis}, {McComas}, {Dayeh},
  {Funsten}, \& {Schwadron}}]{2011ApJ...734....1L}
{Livadiotis}, G., {McComas}, D.~J., {Dayeh}, M.~A., {Funsten}, H.~O., \&
  {Schwadron}, N.~A. 2011, \apj, 734, 1

\bibitem[{{L{\'o}pez-S{\'a}nchez} {et~al.}(2012){L{\'o}pez-S{\'a}nchez},
  {Dopita}, {Kewley}, {Zahid}, {Nicholls}, \&
  {Scharw{\"a}chter}}]{2012MNRAS.426.2630L}
{L{\'o}pez-S{\'a}nchez}, {\'A}.~R., {Dopita}, M.~A., {Kewley}, L.~J., {et~al.}
  2012, \mnras, 426, 2630

\bibitem[{{L{\'o}pez-S{\'a}nchez} \& {Esteban}(2009)}]{2009A&A...508..615L}
{L{\'o}pez-S{\'a}nchez}, {\'A}.~R. \& {Esteban}, C. 2009, \aap, 508, 615

\bibitem[{{L{\'o}pez-S{\'a}nchez} \& {Esteban}(2010)}]{2010A&A...517A..85L}
{L{\'o}pez-S{\'a}nchez}, {\'A}.~R. \& {Esteban}, C. 2010, \aap, 517, A85

\bibitem[{{Mathis} {et~al.}(1977){Mathis}, {Rumpl}, \&
  {Nordsieck}}]{1977ApJ...217..425M}
{Mathis}, J.~S., {Rumpl}, W., \& {Nordsieck}, K.~H. 1977, \apj, 217, 425

\bibitem[{{McLaughlin} \& {Bell}(1998)}]{1998JPhB...31.4317M}
{McLaughlin}, B.~M. \& {Bell}, K.~L. 1998, Journal of Physics B Atomic
  Molecular Physics, 31, 4317

\bibitem[{{Nicholls} {et~al.}(2011){Nicholls}, {Dopita}, {Jerjen}, \&
  {Meurer}}]{2011AJ....142...83N}
{Nicholls}, D.~C., {Dopita}, M.~A., {Jerjen}, H., \& {Meurer}, G.~R. 2011, \aj,
  142, 83

\bibitem[{{Nicholls} {et~al.}(2012){Nicholls}, {Dopita}, \&
  {Sutherland}}]{2012ApJ...752..148N}
{Nicholls}, D.~C., {Dopita}, M.~A., \& {Sutherland}, R.~S. 2012, \apj, 752, 148

\bibitem[{{Nicholls} {et~al.}(2014){Nicholls}, {Dopita}, {Sutherland},
  {Jerjen}, {Kewley}, \& {Basurah}}]{2014ApJ...786..155N}
{Nicholls}, D.~C., {Dopita}, M.~A., {Sutherland}, R.~S., {et~al.} 2014, \apj,
  786, 155

\bibitem[{{Nicholls} {et~al.}(2013){Nicholls}, {Dopita}, {Sutherland},
  {Kewley}, \& {Palay}}]{2013ApJS..207...21N}
{Nicholls}, D.~C., {Dopita}, M.~A., {Sutherland}, R.~S., {Kewley}, L.~J., \&
  {Palay}, E. 2013, \apjs, 207, 21

\bibitem[{{Osterbrock} \& {Ferland}(2006)}]{2006agna.book.....O}
{Osterbrock}, D.~E. \& {Ferland}, G.~J. 2006, {Astrophysics of gaseous nebulae
  and active galactic nuclei}, 2nd edn. (University Science Books)

\bibitem[{{Palay} {et~al.}(2012){Palay}, {Nahar}, {Pradhan}, \&
  {Eissner}}]{2012MNRAS.423L..35P}
{Palay}, E., {Nahar}, S.~N., {Pradhan}, A.~K., \& {Eissner}, W. 2012, \mnras,
  423, L35

\bibitem[{{Pauldrach} {et~al.}(2001){Pauldrach}, {Hoffmann}, \&
  {Lennon}}]{2001A&A...375..161P}
{Pauldrach}, A.~W.~A., {Hoffmann}, T.~L., \& {Lennon}, M. 2001, \aap, 375, 161

\bibitem[{{Peimbert} \& {Costero}(1969)}]{1969BOTT....5....3P}
{Peimbert}, M. \& {Costero}, R. 1969, Boletin de los Observatorios Tonantzintla
  y Tacubaya, 5, 3

\bibitem[{{Pellegrini} {et~al.}(2012){Pellegrini}, {Oey}, {Winkler}, {Points},
  {Smith}, {Jaskot}, \& {Zastrow}}]{2012ApJ...755...40P}
{Pellegrini}, E.~W., {Oey}, M.~S., {Winkler}, P.~F., {et~al.} 2012, \apj, 755,
  40

\bibitem[{{P{\'e}quignot}(2008)}]{2008A&A...478..371P}
{P{\'e}quignot}, D. 2008, \aap, 478, 371

\bibitem[{{P{\'e}quignot} {et~al.}(2001){P{\'e}quignot}, {Ferland}, {Netzer},
  {Kallman}, {Ballantyne}, {Dumont}, {Ercolano}, {Harrington}, {Kraemer},
  {Morisset}, {Nayakshin}, {Rubin}, \& {Sutherland}}]{2001ASPC..247..533P}
{P{\'e}quignot}, D., {Ferland}, G., {Netzer}, H., {et~al.} 2001, in
  Astronomical Society of the Pacific Conference Series, Vol. 247,
  Spectroscopic Challenges of Photoionized Plasmas, ed. G.~{Ferland} \& D.~W.
  {Savin}, 533

\bibitem[{{P{\'e}rez-Montero} {et~al.}(2011){P{\'e}rez-Montero},
  {V{\'{\i}}lchez}, {Cedr{\'e}s}, {H{\"a}gele}, {Moll{\'a}}, {Kehrig},
  {D{\'{\i}}az}, {Garc{\'{\i}}a-Benito}, \&
  {Mart{\'{\i}}n-Gord{\'o}n}}]{2011A&A...532A.141P}
{P{\'e}rez-Montero}, E., {V{\'{\i}}lchez}, J.~M., {Cedr{\'e}s}, B., {et~al.}
  2011, \aap, 532, A141

\bibitem[{{Pradhan} {et~al.}(2006){Pradhan}, {Montenegro}, {Nahar}, \&
  {Eissner}}]{2006MNRAS.366L...6P}
{Pradhan}, A.~K., {Montenegro}, M., {Nahar}, S.~N., \& {Eissner}, W. 2006,
  \mnras, 366, L6

\bibitem[{{Pustilnik} {et~al.}(2004){Pustilnik}, {Kniazev}, {Pramskij},
  {Izotov}, {Foltz}, {Brosch}, {Martin}, \& {Ugryumov}}]{2004A&A...419..469P}
{Pustilnik}, S., {Kniazev}, A., {Pramskij}, A., {et~al.} 2004, \aap, 419, 469

\bibitem[{{R{\'e}my-Ruyer} {et~al.}(2014){R{\'e}my-Ruyer}, {Madden},
  {Galliano}, {Galametz}, {Takeuchi}, {Asano}, {Zhukovska}, {Lebouteiller},
  {Cormier}, {Jones}, {Bocchio}, {Baes}, {Bendo}, {Boquien}, {Boselli},
  {DeLooze}, {Doublier-Pritchard}, {Hughes}, {Karczewski}, \&
  {Spinoglio}}]{2014A&A...563A..31R}
{R{\'e}my-Ruyer}, A., {Madden}, S.~C., {Galliano}, F., {et~al.} 2014, \aap,
  563, A31

\bibitem[{{Seaton}(1975)}]{1975MNRAS.170..475S}
{Seaton}, M.~J. 1975, \mnras, 170, 475

\bibitem[{{Stasi{\'n}ska}(1978)}]{1978A&A....66..257S}
{Stasi{\'n}ska}, G. 1978, \aap, 66, 257

\bibitem[{{Stasi{\'n}ska}(2005)}]{2005A&A...434..507S}
{Stasi{\'n}ska}, G. 2005, \aap, 434, 507

\bibitem[{{Stasi{\'n}ska} {et~al.}(2013){Stasi{\'n}ska}, {Morisset},
  {Sim{\'o}n-D{\'{\i}}az}, {Bresolin}, {Schaerer}, \&
  {Brandl}}]{2013A&A...551A..82S}
{Stasi{\'n}ska}, G., {Morisset}, C., {Sim{\'o}n-D{\'{\i}}az}, S., {et~al.}
  2013, \aap, 551, A82

\bibitem[{{Stasi{\'n}ska} \& {Schaerer}(1999)}]{1999A&A...351...72S}
{Stasi{\'n}ska}, G. \& {Schaerer}, D. 1999, \aap, 351, 72

\bibitem[{{Storey} {et~al.}(2013){Storey}, {Sochi}, \&
  {Badnell}}]{2013arXiv1311.6517S}
{Storey}, P.~J., {Sochi}, T., \& {Badnell}, N.~R. 2013, ArXiv e-prints

\bibitem[{{Tayal}(2007)}]{2007ApJS..171..331T}
{Tayal}, S.~S. 2007, \apjs, 171, 331

\bibitem[{{Townsley} {et~al.}(2003){Townsley}, {Feigelson}, {Montmerle},
  {Broos}, {Chu}, \& {Garmire}}]{2003ApJ...593..874T}
{Townsley}, L.~K., {Feigelson}, E.~D., {Montmerle}, T., {et~al.} 2003, \apj,
  593, 874

\bibitem[{{Tremonti} {et~al.}(2004){Tremonti}, {Heckman}, {Kauffmann},
  {Brinchmann}, {Charlot}, {White}, {Seibert}, {Peng}, {Schlegel}, {Uomoto},
  {Fukugita}, \& {Brinkmann}}]{2004ApJ...613..898T}
{Tremonti}, C.~A., {Heckman}, T.~M., {Kauffmann}, G., {et~al.} 2004, \apj, 613,
  898

\bibitem[{{Tsallis}(2009)}]{2009insm.book.....T}
{Tsallis}, C. 2009, {Introduction to Nonextensive Statistical Mechanics}

\bibitem[{{van Hoof} {et~al.}(2004){van Hoof}, {Weingartner}, {Martin}, {Volk},
  \& {Ferland}}]{2004MNRAS.350.1330V}
{van Hoof}, P.~A.~M., {Weingartner}, J.~C., {Martin}, P.~G., {Volk}, K., \&
  {Ferland}, G.~J. 2004, \mnras, 350, 1330

\bibitem[{{Vilchez} \& {Esteban}(1996)}]{1996MNRAS.280..720V}
{Vilchez}, J.~M. \& {Esteban}, C. 1996, \mnras, 280, 720

\bibitem[{{Yeh} \& {Matzner}(2012)}]{2012ApJ...757..108Y}
{Yeh}, S.~C.~C. \& {Matzner}, C.~D. 2012, \apj, 757, 108

\bibitem[{{York} {et~al.}(2000){York}, {Adelman}, {Anderson}, {Anderson},
  {Annis}, {Bahcall}, {Bakken}, {Barkhouser}, {Bastian}, {Berman}, {Boroski},
  {Bracker}, {Briegel}, {Briggs}, {Brinkmann}, {Brunner}, {Burles}, {Carey},
  {Carr}, {Castander}, {Chen}, {Colestock}, {Connolly}, {Crocker}, {Csabai},
  {Czarapata}, {Davis}, {Doi}, {Dombeck}, {Eisenstein}, {Ellman}, {Elms},
  {Evans}, {Fan}, {Federwitz}, {Fiscelli}, {Friedman}, {Frieman}, {Fukugita},
  {Gillespie}, {Gunn}, {Gurbani}, {de Haas}, {Haldeman}, {Harris}, {Hayes},
  {Heckman}, {Hennessy}, {Hindsley}, {Holm}, {Holmgren}, {Huang}, {Hull},
  {Husby}, {Ichikawa}, {Ichikawa}, {Ivezi{\'c}}, {Kent}, {Kim}, {Kinney},
  {Klaene}, {Kleinman}, {Kleinman}, {Knapp}, {Korienek}, {Kron}, {Kunszt},
  {Lamb}, {Lee}, {Leger}, {Limmongkol}, {Lindenmeyer}, {Long}, {Loomis},
  {Loveday}, {Lucinio}, {Lupton}, {MacKinnon}, {Mannery}, {Mantsch}, {Margon},
  {McGehee}, {McKay}, {Meiksin}, {Merelli}, {Monet}, {Munn}, {Narayanan},
  {Nash}, {Neilsen}, {Neswold}, {Newberg}, {Nichol}, {Nicinski}, {Nonino},
  {Okada}, {Okamura}, {Ostriker}, {Owen}, {Pauls}, {Peoples}, {Peterson},
  {Petravick}, {Pier}, {Pope}, {Pordes}, {Prosapio}, {Rechenmacher}, {Quinn},
  {Richards}, {Richmond}, {Rivetta}, {Rockosi}, {Ruthmansdorfer}, {Sandford},
  {Schlegel}, {Schneider}, {Sekiguchi}, {Sergey}, {Shimasaku}, {Siegmund},
  {Smee}, {Smith}, {Snedden}, {Stone}, {Stoughton}, {Strauss}, {Stubbs},
  {SubbaRao}, {Szalay}, {Szapudi}, {Szokoly}, {Thakar}, {Tremonti}, {Tucker},
  {Uomoto}, {Vanden Berk}, {Vogeley}, {Waddell}, {Wang}, {Watanabe},
  {Weinberg}, {Yanny}, \& {Yasuda}}]{2000AJ....120.1579Y}
{York}, D.~G., {Adelman}, J., {Anderson}, Jr., J.~E., {et~al.} 2000, \aj, 120,
  1579

\bibitem[{{Zastrow} {et~al.}(2013){Zastrow}, {Oey}, {Veilleux}, \&
  {McDonald}}]{2013ApJ...779...76Z}
{Zastrow}, J., {Oey}, M.~S., {Veilleux}, S., \& {McDonald}, M. 2013, \apj, 779,
  76

\end{thebibliography}

\newpage
\section{Appendix: Revisions to the Mappings photoionization modeling code}

The current version of the Mappings photoionization modeling code is IV.1.4. As the work described in this paper depends heavily on the Mappings code, it is useful to summarize the major changes since the previous version (IV.0) \citep{2013ApJS..208...10D}. A new integration scheme has been implemented, along with improved equilibrium calculations, to achieve a more precise balance of heating and cooling. The new code model performs better in \HII regions and planetary nebula models when compared to the Lexington benchmarks \citep{2001ASPC..247..533P}, with more consistent ionization state and temperature averages, compared to previous Mappings versions. Photoelectric cross sections for C, N, O, He~{\sc i} and He~{\sc ii} have been revised to take into account the latest experimental and model atomic data. Hydrogen and helium charge exchange reactions for both recombination and ionization have been updated, as are the ways in which these are handled. The hydrogenic free-bound recombination calculations have been revised, with new coefficients, to ensure detailed balance and energy conservation.

\end{document}